\PassOptionsToPackage{square,comma,numbers,sort&compress,super}{natbib}
\documentclass[aps,prb,twocolumn,superscriptaddress]{revtex4-1}

\usepackage{graphicx}
\usepackage[font={small}]{caption}
\usepackage{subcaption}
\usepackage{multirow}
\usepackage[dvipsnames]{xcolor}
\usepackage{colortbl}
\usepackage{mathtools}
\usepackage{adjustbox}
\usepackage{amsmath}
\usepackage{amssymb}
\usepackage{listings}
\usepackage{multirow}
\usepackage{float}
\usepackage{url}
\usepackage[normalem]{ulem}

\usepackage{natbib}
\usepackage{achemso}
\setkeys{acs}{maxauthors = 0}

\begin{document}

\setlength{\tabcolsep}{4pt} 
\renewcommand{\arraystretch}{1.05}
\renewcommand{\thetable}{\arabic{table}}

\title{Enhanced Twist-Averaging Technique for Magnetic Metals: Applications using Quantum Monte Carlo}

\author{Abdulgani Annaberdiyev}
\affiliation{Center for Nanophase Materials Sciences, Oak Ridge National Laboratory, Oak Ridge, Tennessee 37831, USA}
\email{annaberdiyea@ornl.gov}

\author{Panchapakesan Ganesh}
\affiliation{Center for Nanophase Materials Sciences, Oak Ridge National Laboratory, Oak Ridge, Tennessee 37831, USA}

\author{Jaron T. Krogel}
\affiliation{Materials Science and Technology Division, Oak Ridge National Laboratory, Oak Ridge, Tennessee 37831, USA}

\begin{abstract}

We propose an improved twist-averaging scheme for quantum Monte Carlo methods that use converged Kohn-Sham or Hartree-Fock orbitals as the reference.
This twist-averaging technique is tailored to sample the Brillouin zone of magnetic metals, although it naturally extends to nonmagnetic conducting systems.
The proposed scheme aims to reproduce the reference magnetization and achieves charge neutrality by construction, thus avoiding the large energy fluctuations and the postprocessing needed to correct the energies.
It shows the most robust convergence of total energy and magnetism to the thermodynamic limit when compared to four other twist-averaging schemes.
Diffusion Monte Carlo applications are shown on nonmagnetic Al and ferromagnetic $\alpha$-Fe.
The cohesive energy of Al in the thermodynamic limit shows an excellent agreement with the experimental result.
Furthermore, the magnetic moments in $\alpha$-Fe exhibit rapid convergence with an increasing number of twists.

\end{abstract}

\maketitle

\begin{figure}[!htbp]
\centering
\includegraphics[width=0.4\textwidth]{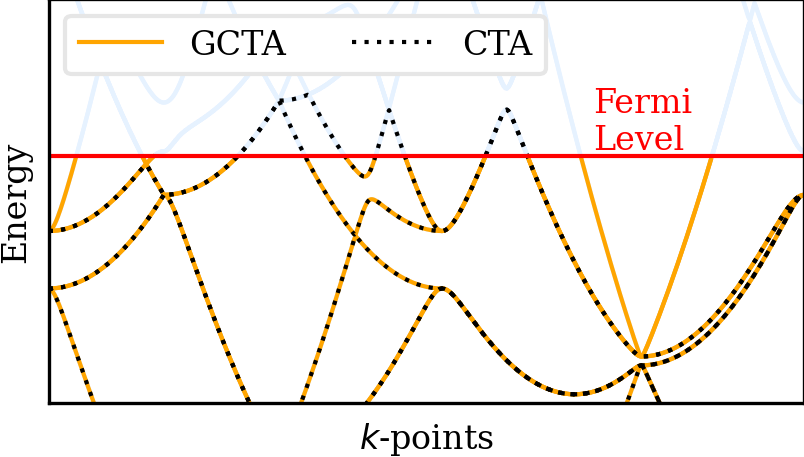}
\label{fig:abstract_graphics}
\end{figure}

\section{Introduction}
\label{sec:intro}

Accurate characterization of magnetic metals in quantum Monte Carlo (QMC) \cite{foulkes_quantum_2001, kolorenc_applications_2011, hunt_quantum_2018, austin_quantum_2012, luchow_quantum_2011, acioli_review_1997} methods poses significant challenges.
One aspect of the problem is related to the $k$-point sampling of the Brillouin zone (BZ) in metals.
In contrast to insulators, the difficulty in $k$-point sampling in metals arises due to bands crossing the Fermi level (FL), which results in varying numbers of occupations in different $k$-points.
Therefore, this fluctuation in the occupation numbers in reciprocal space must be appropriately modeled to enable high-accuracy calculations~\cite{lin_twist-averaged_2001}.

In QMC methods, $k$-point sampling is accounted for by using twisted-boundary conditions in the simulation cell.
This is referred to as twist-averaging (TA), which is unavoidable in metallic systems to achieve practical results \cite{hood_diffusion_2012, holzmann_theory_2016}.
The cost to reach a target statistical accuracy in QMC nominally does not depend on the size of the $k$-point grid, $\mathcal{O}(1)$, due to statistical averaging of the total sample size, which can be kept constant.
However, practical limitations can burden the dense $k$-point sampling in QMC.
For example, each $k$-point is not trivially parallelizable due to costly warmup/equilibration period in time propagation \cite{lin_twist-averaged_2001, dagrada_exact_2016}, which can be especially significant in diffusion Monte Carlo.
Another complication is the large disc storage required when each $k$-point accumulates a separate real-space grid observable, such as charge or spin density \cite{drummond_finite-size_2008}.
The large files generated can make data analysis slow and cumbersome.
Therefore, obtaining fast and robust convergence with respect to the $k$-grid is desirable.

Specifically, we do not study \textit{which} $k$-points give the best convergence to the thermodynamic limit (TDL) \cite{rajagopal_quantum_1994, rajagopal_variational_1995, drummond_finite-size_2008, dagrada_exact_2016}; rather, the focus is on \textit{how} a given a set of $k$-points should be occupied within a twist-averaging scheme to achieve rapid convergence in the total energy and magnetic moment with the supercell size.
The chosen $k$-point occupations should also ideally lead to linear behavior in QMC total energy versus the inverse electron count to facilitate extrapolation to the TDL \cite{kolorenc_applications_2011}.
This is desired since perfect linear convergence would make it sufficient to use only the smallest supercells to achieve minimum bias, thus avoiding the computationally intensive large supercell calculations that are also sometimes impossible in practice due to random-access memory limitations in the central/graphics processing unit \cite{ali_chapter_2013, luo_efficient_2018}.
In other words, the ideal occupation scheme should yield highly predictable extrapolation behavior.

The traditional practice of averaging twisted boundary conditions in real-space QMC has been a widely adopted methodology for over two decades \cite{lin_twist-averaged_2001}.
Various strategies for selecting twists to be incorporated into the averaging process and determining their occupancy have been proposed. While the Monkhorst-Pack grid \cite{monkhorst_special_1976}, representing a uniform $k$-point set, is prevalent \cite{lin_twist-averaged_2001, drummond_finite-size_2008, qin_benchmark_2016, clay_benchmarking_2016, azadi_efficient_2019, azadi_low-density_2022}, alternative methods for $k$-point selection, such as pseudorandom \cite{lin_twist-averaged_2001, drummond_finite-size_2008, qin_benchmark_2016, azadi_efficient_2019}, quasi-random \cite{lin_twist-averaged_2001, qin_benchmark_2016}, dynamic \cite{lin_twist-averaged_2001}, and patch/pocket reweighting approaches \cite{holzmann_theory_2016}, have been suggested.
The advent of high-performance computing hardware \cite{ali_chapter_2013} has enabled denser choices for the integration grid, establishing the use of a uniform grid as the predominant method in QMC.
Nonetheless, the question of how the chosen set of $k$-points should be occupied in metallic systems persists without a definitive, standardized approach.

A widely embraced approach involves canonical twist-averaging (CTA) \cite{lin_twist-averaged_2001, zong_quantum_2002, drummond_finite-size_2008, clay_benchmarking_2016, azadi_low-density_2022}, wherein the lowest energy single-particle states per super twist are occupied until charge neutrality is achieved for each super twist.
In this scheme, each super twist maintains a fixed charge (zero) but exhibits a fluctuating magnetic moment if present.
Notably, each super twist wave function shares the same eigenvalue of the total number operator while displaying different eigenvalues of $\hat{S}_z$, or, equivalently, different spin-resolved number operators.
Although CTA has proven successful in insulators, its application to metallic systems has encountered challenges as it fails to accurately represent the momentum distribution around the Fermi surface \cite{lin_twist-averaged_2001}.
This is because CTA may occupy single-particle states above the Fermi level (or may leave states below the Fermi level unoccupied) to achieve charge neutrality for each twist.

To address this limitation, the grand-canonical twist-averaging (GCTA) method was proposed early on \cite{lin_twist-averaged_2001}, aiming to occupy single-particle levels in each super twist strictly based on the Fermi level obtained in the thermodynamic limit (TDL).
Consequently, in this conventional form, individual super twists exhibit variable charge, with charge neutrality only achieved in the TDL.
However, the nonzero net charge associated with conventional GCTA can result in erratic energy and charge density fluctuations, hindering its widespread adoption in practical material QMC calculations \cite{lin_twist-averaged_2001}.
So far, it has only been used in a few systems, such as the homogeneous electron gas (HEG) \cite{lin_twist-averaged_2001, zong_quantum_2002, drummond_finite-size_2008, holzmann_renormalization_2009}, Hubbard model \cite{qin_benchmark_2016}, hydrogen and helium \cite{holzmann_renormalization_2009}.

A more recent refinement to the conventional GCTA approach involves postcorrecting energies based on the grand potential to achieve smooth convergence \cite{azadi_efficient_2019}.
Encouraging results have been observed for the GCTA of grand potential, particularly in total energies and their components.
Our work aims to enhance the conventional GCTA in a manner that eliminates reliance on postcorrections and extends its applicability beyond total energies and its components.
The objective is to devise a GCTA method that demonstrates smooth convergence across a spectrum of observables, including charge density, spin density (i.e., magnetization), momentum distribution, and other measurable parameters.
This research focuses on addressing this gap and demonstrates, in subsequent sections, that the proposed modification to conventional GCTA indeed results in smooth convergence across various quantities, including total, kinetic, and potential energies, cell magnetization, atomic moments, and charge densities.

The paper is organized as follows.
In the next Section (\ref{sec:theory}), we describe the terminology and the theory of different twist-averaging schemes and provide the computational details.
Then, in Section \ref{sec:results}, we apply and compare these schemes in selected systems.
We discuss the results in Section \ref{sec:discussion}, and we make concluding remarks in Section \ref{sec:conclusions}.

\section{Theory and Methodology}
\label{sec:theory}

Under twisted boundary conditions, the many-body wave function $\Psi_{\mathbf{k}_\mathrm{s}}$ at a twist wave vector $\mathbf{k}_\mathrm{s}$ picks up a phase as an electron is translated by a supercell lattice vector $\mathbf{R}_\mathrm{s}$:
\begin{multline}
    \Psi_{\mathbf{k}_\mathrm{s}}(\mathbf{r}_1, ..., \mathbf{r}_n + \mathbf{R}_\mathrm{s}, ..., \mathbf{r}_{N^\mathrm{s}_e}) = \\
    = \mathrm{exp}(i \mathbf{k}_\mathrm{s} \cdot \mathbf{R}_\mathrm{s}) \Psi_{\mathbf{k}_\mathrm{s}}(\mathbf{r}_1, ..., \mathbf{r}_n, ..., \mathbf{r}_{N^\mathrm{s}_e})
\end{multline}
where $N^\mathrm{s}_e$ is the number of electrons in the supercell and $\theta = \mathbf{k}_\mathrm{s} \cdot \mathbf{R}_\mathrm{s}$ is called the twist angle.
Throughout this work, we use the term ``twist-mesh" when referring to QMC calculations.
The observables are obtained by averaging over a chosen set of twists:
\begin{equation}
    \left< \hat{O} \right> = \frac{1}{Z_{\theta}} \sum_{\mathbf{k}_\mathrm{s}} \left< \Psi_{\mathbf{k}_\mathrm{s}} \left| \hat{O} \right| \Psi_{\mathbf{k}_\mathrm{s}} \right>
\end{equation}
where $Z_{\theta}=N^3_{\theta}$ is the total number of twists included in the twist-averaging for a given supercell.
This is similar to the $k$-point averaging of BZ carried out in density functional theory (DFT) calculations.
However, the role of $k$-points in QMC is introduced as a supercell boundary condition using phase angle $\theta$.
For example, in $\Gamma$-point calculations, $\mathbf{k}_\mathrm{s} = 0$ and $\theta = 0$, resulting in periodic boundary conditions (PBCs).
For $\theta = \pi$, the many-body wave function picks up a negative sign as an electron wraps around the simulation cell and returns to its initial position (antiperiodic boundary conditions).
In general, $\theta$ can take any value between $-\pi$ and $\pi$ (twisted boundary conditions).

Similarly, $k$-point sampling in QMC (and DFT) can also be accounted for by using progressively larger supercells at a single twist.
In this case, the $k$-points in the primitive cell fold into the $k$-points of the supercell, where the reciprocal space is now reduced.
For example, using a single $\Gamma$-point calculation, namely, PBC, the $[2\times2\times2]$ supercell of a simple cubic system will include the center, face, edge, and corner ($\Gamma$, X, M, R) high-symmetry points of the BZ.
In the limit of an infinitely large supercell at the $\Gamma$-point, one samples all $k$-points in the BZ (perfect integration) as all primitive cell $k$-points fold into the $\Gamma$-twist.
In other words, the reciprocal space shrinks to a single point, $\Gamma$-point, which can be sampled using only that single twist.

\subsection{Twist-Averaging Occupation Methods}
\label{sec:ta_methods}

Let $N^\mathrm{p}_e$ be the number of electrons in a neutral primitive cell, and $M^\mathrm{p}_{\mathrm{DFT}} \in \mathbb{R}$ is the corresponding cell magnetization obtained in DFT.
In QMC, we use uniformly-tiled supercells as $Z_T = [T \times T \times T]$ from the primitive cell.
In the limit of $Z_T\rightarrow \infty$, we would like to achieve the following conditions for the supercell occupations of spin up $N^\mathrm{s}_{\uparrow}$ and spin down $N^\mathrm{s}_{\downarrow}$ at the $\Gamma$-twist:
\begin{equation}
    \label{eq:cond_charge}
    N^\mathrm{s}_{\uparrow} + N^\mathrm{s}_{\downarrow} = N^\mathrm{p}_e \cdot Z_T \quad\quad Z_T = T^3,
\end{equation}
\begin{equation}
    \label{eq:cond_moment}
    N^\mathrm{s}_{\uparrow} - N^\mathrm{s}_{\downarrow} = M^\mathrm{p}_{\mathrm{DFT}} \cdot Z_T \quad\quad M^\mathrm{p}_{\mathrm{DFT}} \in \mathbb{R}.
\end{equation}
The equality in Equation \ref{eq:cond_moment} holds since any real number can be expressed as a fraction of arbitrarily large numbers.
In practice, $Z_T$ is usually too small to achieve accurate BZ sampling, and thus, twist-averaging needs to be introduced.
Consider a supercell with a twist-mesh size of $Z_\theta$.
For each supercell twist $\theta$, the corresponding spin up/down occupations $N^\mathrm{s}_{\uparrow}(\theta)$/$N^\mathrm{s}_{\downarrow}(\theta)$ can be set using various approaches.
The key quantity by which the TA methods are classified is the charge in each twist:
\begin{equation}
    q(\theta) = N^\mathrm{s}_{\uparrow}(\theta) + N^\mathrm{s}_{\downarrow}(\theta) - (N^\mathrm{p}_e \cdot Z_T).
\end{equation}
In the canonical twist-averaging (CTA) methods, the charge is constant for every twist and equals zero, $q(\theta) = 0$.
In the so-called grand-canonical twist-averaging (GCTA) methods, the charge is allowed to vary as a function of twist: $q(\theta) \neq 0$.
Despite its name, the total number of particles in GCTA is constant during the QMC time propagation for each twist.

We assume that well-converged Kohn-Sham (KS) or Hartree-Fock (HF) orbitals are available to the QMC method.
These orbitals are provided to the single-reference QMC to be occupied by the following schemes.

\subsubsection{GCTA-DFT}
This subsection details the grand-canonical twist-averaging using the DFT Fermi level (which we denote as GCTA-DFT here).
In this conventional approach, the orbital is occupied if the KS eigenvalue ($\varepsilon^{\uparrow/\downarrow}_i$) is lower than the DFT-determined Fermi level ($E^{\mathrm{DFT}}_\mathrm{F}$):
\begin{equation}
    \label{eqn:n_gcta}
    N^\mathrm{s}_{\uparrow/\downarrow}({\theta}) = \sum^{N^\mathrm{p}_{\mathrm{KS}}}_i{f_{\uparrow/\downarrow}(i)}
\end{equation}
where $N^\mathrm{p}_{\mathrm{KS}}$ is the number of KS orbitals solved in DFT for each channel.
$f_{\uparrow/\downarrow}(i)$ is defined as:
\begin{equation}
    \label{eqn:f_gcta}
    f_{\uparrow/\downarrow}(i)= 
    \begin{cases}
        1, & \text{if } (\varepsilon^{\uparrow/\downarrow}_i < E_\mathrm{F}) \\
        0, & \text{otherwise}.
    \end{cases}
\end{equation}
In this case, the DFT Fermi level determined using the full DFT $k$-mesh is employed in the QMC:
\begin{equation}
    E_\mathrm{F} = E^{\mathrm{DFT}}_\mathrm{F}.
\end{equation}
The resulting twist-averaged system in QMC is not guaranteed to be charge-neutral using GCTA-DFT:
\begin{equation}
    Q_{\mathrm{TA}} =  \frac{1}{Z_{\theta}} \sum^{Z_{\theta}}_{\theta} q(\theta) \neq 0.
\end{equation}
The reason for this is that $E^{\mathrm{DFT}}_\mathrm{F}$ is determined within the full $k$-mesh in DFT, and coarser $k$-mesh calculations might need a shift in the Fermi level to achieve charge-neutrality (the strategy of the next scheme).
A secondary reason exists for $Q_{\mathrm{TA}} \neq 0$ even when the full DFT $k$-mesh is employed in QMC.
Charge smearing is often used in DFT for converging the metallic systems \cite{methfessel_high-precision_1989, marzari_thermal_1999}, resulting in fractional occupations near $E_\mathrm{F}$.
However, these fractional occupations get rounded in the QMC, which might slightly modify the $E_\mathrm{F}$ needed for charge-neutrality.
The proposed improvements will fix both of the above sources of issues.

\subsubsection{GCTA-AFL}
This subsection introduces GCTA with an adapted Fermi level (GCTA-AFL).
In this improved scheme, the orbitals are again occupied as given in Equations \ref{eqn:n_gcta} and \ref{eqn:f_gcta}, but the Fermi level is shifted so that the twist-averaged system is charge neutral for any value of $Z_\theta$, $E_\mathrm{F} = E^{\mathrm{AFL}}_\mathrm{F}$.
To determine $E^{\mathrm{AFL}}_\mathrm{F}$, consider a set $\mathcal{E}$ which contains the sorted list of \textit{all} KS eigenvalues of a supercell characterized by $Z_T$ and $Z_\theta$:
\begin{equation}
    \label{eqn:eig_afl_set}
    \mathcal{E}_i = \left\{\varepsilon^\chi_1, \varepsilon^\chi_2, \cdots , \varepsilon^\chi_i, \cdots, \varepsilon^\chi_L \right\} \quad (\forall i \quad \varepsilon^\chi_i < \varepsilon^\chi_{i+1})
\end{equation}
where $\chi = \left\{ \uparrow , \downarrow \right\}$ contains both spin channels, and $\mathcal{E}$ has a length of $L = 2 \cdot N^\mathrm{p}_\mathrm{KS} \cdot Z_T\cdot Z_{\theta}$.
Now consider the index $\lambda$:
\begin{equation}
    \lambda = N^\mathrm{p}_e \cdot Z_T\cdot Z_{\theta}
\end{equation}
which is obtained using electron counting.
Now, $\lambda$ can be used to set the $E_\mathrm{F}$:
\begin{equation}
    E^{\mathrm{AFL}}_\mathrm{F} = \frac{\mathcal{E}_\lambda + \mathcal{E}_{\lambda+1}}{2}.
\end{equation}
Within the mean-field framework, this is equivalent to occupying the $\Gamma$-point of a charge-neutral supercell that has a $Z_T\cdot Z_{\theta}$ number of primitive cells.  
Namely, the result is $Q_{\mathrm{TA}} = 0$ ($q(\theta) \neq 0$) by definition, at the price of a fluctuating $E_\mathrm{F}$ as $Z_T$ or $Z_\theta$ is changed.
In addition, $E_\mathrm{F}$ is set globally for both spin channels, which can introduce a small deviation to the reference magnetization $M^\mathrm{p}_{\mathrm{DFT}}$.

\subsubsection{GCTA-SAFL}
This subsection introduces GCTA with a spin-adapted Fermi level (GCTA-SAFL).
This is the scheme that we propose as the most accurate technique.
Here, the Fermi levels for up and down channels are shifted independently to target the reference cell magnetization, which is the DFT magnetization ($M^\mathrm{p}_{\mathrm{DFT}}$) throughout this work.
The KS eigenvalues are now sorted separately for each spin channel:
\begin{equation}
    \label{eqn:safl_set_up}
    \mathcal{E}_i^\uparrow = \left\{\varepsilon^\uparrow_1, \varepsilon^\uparrow_2, \cdots , \varepsilon^\uparrow_i, \cdots, \varepsilon^\uparrow_\ell \right\} \quad (\forall i \quad \varepsilon^\uparrow_i < \varepsilon^\uparrow_{i+1})
\end{equation}
\begin{equation}
    \label{eqn:safl_set_dn}
    \mathcal{E}_i^\downarrow = \left\{\varepsilon^\downarrow_1, \varepsilon^\downarrow_2, \cdots , \varepsilon^\downarrow_i, \cdots, \varepsilon^\downarrow_\ell \right\} \quad (\forall i \quad \varepsilon^\downarrow_i < \varepsilon^\downarrow_{i+1})
\end{equation}
where the length of the lists is $\ell = N^\mathrm{p}_\mathrm{KS} \cdot Z_T\cdot Z_{\theta}$.
Then, the following up ($u$) and down ($d$) indices can be used to set the $E_\mathrm{F}$: 
\begin{equation}
    u = \mathrm{Round}\left[ \frac{(N^\mathrm{p}_e + M^\mathrm{p}_{\mathrm{DFT}}) \cdot Z_T\cdot Z_{\theta}}{2} \right]
\end{equation}
\begin{equation}
    d = (N^\mathrm{p}_e \cdot Z_T\cdot Z_{\theta}) - u
\end{equation}
\begin{equation}
    E^{\mathrm{SAFL-\uparrow}}_\mathrm{F} = \frac{\mathcal{E}^{\uparrow}_u + \mathcal{E}^{\uparrow}_{u+1}}{2} \quad\quad E^{\mathrm{SAFL-\downarrow}}_\mathrm{F} = \frac{\mathcal{E}^{\downarrow}_d + \mathcal{E}^{\downarrow}_{d+1}}{2}
\end{equation}
The $u$ and $d$ indices are again derived by electron counting; see Equations \ref{eq:cond_charge}, \ref{eq:cond_moment}.
This occupation scheme again results in $Q_{\mathrm{TA}} = 0$ but also closely follows the reference magnetization ($M^\mathrm{p}_{\mathrm{DFT}}$).
Effectively, some KS eigenvalues in one spin channel that are higher in energy than the other spin channel eigenvalues are allowed to be occupied.
As shown later, this small cost is well worth paying to achieve robust convergence in magnetism.
For non-magnetic (NM) or antiferromagnetic (AFM) systems where up and down eigenvalues are the same, GCTA-AFL and GCTA-SAFL schemes become equivalent.

\subsubsection{CTA-DFT}
This subsection details canonical twist-averaging using the lowest DFT eigenvalues (which we denote as CTA-DFT here).
In this scheme, each twist is charge-neutral, $q(\theta) = 0$.
However, $N^\mathrm{s}_{\uparrow}({\theta})$ and $N^\mathrm{s}_{\downarrow}({\theta})$ are not explicitly set to reproduce the DFT magnetization ($M^\mathrm{p}_\mathrm{DFT}$).
Rather, the occupations are solely decided by using the KS eigenvalues, occupying the lowest $\varepsilon_i$ values from both channels until charge-neutrality is met.
To determine the occupations, consider a set $\mathcal{E}(\theta)$ which contains the sorted list of \textit{all} eigenvalues at a supercell twist $\theta$:
\begin{equation}
    \mathcal{E}(\theta)_i = \left\{\varepsilon^\chi_1, \varepsilon^\chi_2, \cdots , \varepsilon^\chi_i, \cdots, \varepsilon^\chi_L \right\} \quad (\forall i \quad \varepsilon^\chi_i < \varepsilon^\chi_{i+1})
\end{equation}
where $\chi = \left\{ \uparrow , \downarrow \right\}$ contains both spin channels, and $\mathcal{E}(\theta)$ has a length of $L = 2 \cdot N^\mathrm{p}_\mathrm{KS} \cdot Z_T$.
Now consider the index $\lambda$:
\begin{equation}
    \lambda = N^\mathrm{p}_e \cdot Z_T
\end{equation}
which can be used to set the Fermi level:
\begin{equation}
    E(\theta)^{\mathrm{CTA-DFT}}_\mathrm{F} = \frac{\mathcal{E}(\theta)_\lambda + \mathcal{E}(\theta)_{\lambda+1}}{2}.
\end{equation}
Note that $E_\mathrm{F}$ is now a function of twist $\theta$.
The occupations are again set using Equations \ref{eqn:n_gcta} and \ref{eqn:f_gcta}, although there is no variation in $q(\theta)$ ($q(\theta) = 0$, $Q_{\mathrm{TA}} = 0$), which is achieved by allowing $E_\mathrm{F}$ to vary per twist.

\subsubsection{CTA-INS}
This subsection details the CTA used in insulators (which we denote as CTA-INS here).
In this procedure, each twist $\theta$ is charge neutral and has the same magnetization:
\begin{equation}
    M^\mathrm{s}(\theta) = N^\mathrm{s}_{\uparrow}(\theta) - N^\mathrm{s}_{\downarrow}(\theta) = \text{Constant}.
\end{equation}
As this is the scheme used in insulators, there is an additional constraint: for an even/odd number of total electrons $N^\mathrm{s}_e(\theta)$, the magnetization $M^\mathrm{s}(\theta)$ should also be even/odd, respectively.
Therefore, we introduce the function:
\begin{equation}
    \mathcal{F}(x, y)= 
    \begin{cases}
        2 \cdot \mathrm{Round}\left[ x / 2 \right], & \text{if } (y \text{ is even}) \\
        2 \cdot \mathrm{Floor}\lfloor x / 2 \rfloor + 1, & \text{if } (y \text{ is odd}) \\
    \end{cases}
\end{equation}
which can then be used to set the occupations as:
\begin{equation}
    N^\mathrm{s}_{\uparrow}({\theta}) = \frac{N^\mathrm{p}_e \cdot Z_T+ \mathcal{F}(M^\mathrm{p}_\mathrm{DFT} \cdot Z_T, N^\mathrm{p}_e \cdot Z_T))}{2}
\end{equation}
\begin{equation}
    N^\mathrm{s}_{\downarrow}({\theta}) = N^\mathrm{p}_e \cdot Z_T- N^\mathrm{s}_{\uparrow}({\theta}).
\end{equation}
This is the occupation scheme with the most strict constraints as both the charge $q(\theta) = 0$ and magnetization $M^\mathrm{s}(\theta)$ are not allowed to vary. 
Nevertheless, we include it in the comparisons for completeness.
For NM and AFM systems with an even number of electrons, CTA-DFT and CTA-INS schemes become equivalent.

\subsection{Systems}
\label{sec:systems}


We decided to study two distinct systems to benchmark the occupation schemes.
One is the main group elemental solid, Al, which is a paramagnetic metal \cite{young_university_2011} (henceforth referred to as nonmagnetic).
It has a face-centered-cubic (fcc) structure with a conventional cell consisting of 4 Al atoms \cite{nakashima_crystallography_2018}.
Al was chosen since it has a highly dispersive band structure (see Supplementary Information (SI)), which makes it an ideal system for studying various twist-averaging occupation schemes.
The other system is a transition metal elemental solid, $\alpha$-Fe, which is a ferromagnetic (FM) conductor \cite{chen_experimental_1995}.
The $\alpha$ phase is the stable phase of Fe at low temperatures and pressures \cite{boehler_high-pressure_2000}, and it has a body-centered-cubic (bcc) structure with a conventional cell consisting of 2 Fe atoms.
$\alpha$-Fe serves as a test case that is more complicated than that of a main group element and provides an understanding of the convergence of magnetism for various occupation schemes.
We have used experimental geometries extrapolated to $T = 0$~K temperature for both Al \cite{nakashima_crystallography_2018} ($a = 4.0317(2)$~\AA) and $\alpha$-Fe \cite{owen_low-temperature_1954} ($a = 2.8598$~\AA).

\subsection{Computational Details}
\label{sec:computation}

For QMC calculations, we chose a $\Gamma$-centered mesh with an even grid in every direction:
\begin{equation}
    {Z}_\theta = N_{\theta}^3 \quad\quad N_{\theta} \in \mathrm{even}.
\end{equation}
The even grid was chosen to consistently include the center, face, edge, and corner high-symmetry points ($\Gamma$, X, M, and R) in the BZ of the smallest considered cell; see the SI for locations of these points.
This grid type allowed us to tile the required supercells ($[2 \times 2 \times 2]$ and $[3 \times 3 \times 3]$) and showed good convergence in Al and $\alpha$-Fe.
The chosen grid was kept fixed for various TA schemes to reveal the differences due to occupations.
The smallest simulated cells are the fcc conventional cell for Al (4 atoms) and the bcc conventional cell for $\alpha$-Fe (2 atoms); see the SI for the figures.
These cells were taken as primitive cells for tiling larger supercells in QMC calculations.

The Al and Fe atoms were represented by correlation-consistent effective core potentials (ccECPs \cite{bennett_new_2018} for Al and ccECP-soft \cite{kincaid_correlation_2022} for Fe).
Bulk self-consistent field (SCF) calculations were carried out using \texttt{Quantum Espresso} \cite{giannozzi_quantum_2009, giannozzi_advanced_2017, giannozzi_quantum_2020}, while the atomic SCF calculations employed \texttt{PySCF} \cite{sun_pyscf_2018, sun_recent_2020}.
In \texttt{Quantum Espresso}, the KS orbitals were converged at $[24 \times 24 \times 24]$ $k$-mesh in the BZ of the smallest cell (1183 $k$-points in the irreducible BZ) using 400 Ry kinetic energy cutoff; see the SI for convergence tests.
LDA exchange-correlation functional \cite{perdew_self-interaction_1981} was used for bulk Al, while LDA$+U$(5.5~eV) \cite{cococcioni_linear_2005, himmetoglu_hubbard-corrected_2014} was used for $\alpha$-Fe.
We used LDA$+U$ in $\alpha$-Fe to enhance magnetism and amplify the differences between various TA schemes.
Therefore, we do not expect the Fe atomic moments to agree with the experiments.

All QMC calculations were carried out using \texttt{QMCPACK} \cite{kim_qmcpack_2018, kent_qmcpack_2020}.
We used single-reference Slater-Jastrow-type (SJ) trial wave functions in QMC.
Specifically, we used the real-space fixed-node/fixed-phase diffusion Monte Carlo (DMC) to evaluate the observables.
In the following, we only report the raw energies and do not apply any postprocessing to correct the energies \cite{chiesa_finite-size_2006}.
Also, the electron-electron interactions use the standard Ewald potential \cite{ewald_berechnung_1921, foulkes_quantum_2001, drummond_finite-size_2008}, and other types of interactions, such as Yukawa \cite{salin_ewald_2000} or MPC \cite{fraser_finite-size_1996, williamson_elimination_1997, kent_finite-size_1999}, are not considered.
Throughout this work, the DMC time step was set to $\Delta \tau = 0.005$~Ha$^{-1}$.
Size-consistent T-moves \cite{casula_size-consistent_2010} were used in sampling the nonlocal pseudopotentials.
QMC up and down spin channel densities were accumulated as a grid histogram with $[100 \times 100 \times 100]$ points per primitive cell.
The atomic moments and charges were calculated using densities within a spherical radius $R_\mathrm{c}$, which was chosen as the midpoint between the nearest-neighbor atoms.
The extrapolated DMC moments and charges were obtained using the mixed DMC estimator and variational Monte Carlo (VMC) estimator \cite{foulkes_quantum_2001}.
All Jastrow factors were optimized for energy at the $k=\Gamma$ twist and reused at other twists.
See the SI for further details on the QMC methodology.
Some images in the SI were generated using \texttt{XCrySDen} software \cite{kokalj_xcrysdennew_1999}.

\section{Results}
\label{sec:results}

Subsections \ref{sec:twist_conv} and \ref{sec:supercell} present the results from applying the TA schemes described in subsection \ref{sec:ta_methods} to Al and $\alpha$-Fe.
First, in subsection \ref{sec:twist_conv}, we consider the convergence of physical properties in the smallest cell ($Z_T = 1$) by increasing only the twist-mesh $Z_\theta$.
Next, in subsection \ref{sec:supercell}, we study the convergence with respect to supercell size $Z_T$.

\subsection{Twist-Mesh Size Convergence}
\label{sec:twist_conv}

\subsubsection{Al}

\begin{figure*}[!htbp]
\centering
\begin{subfigure}{0.33\textwidth}
\includegraphics[width=\textwidth]{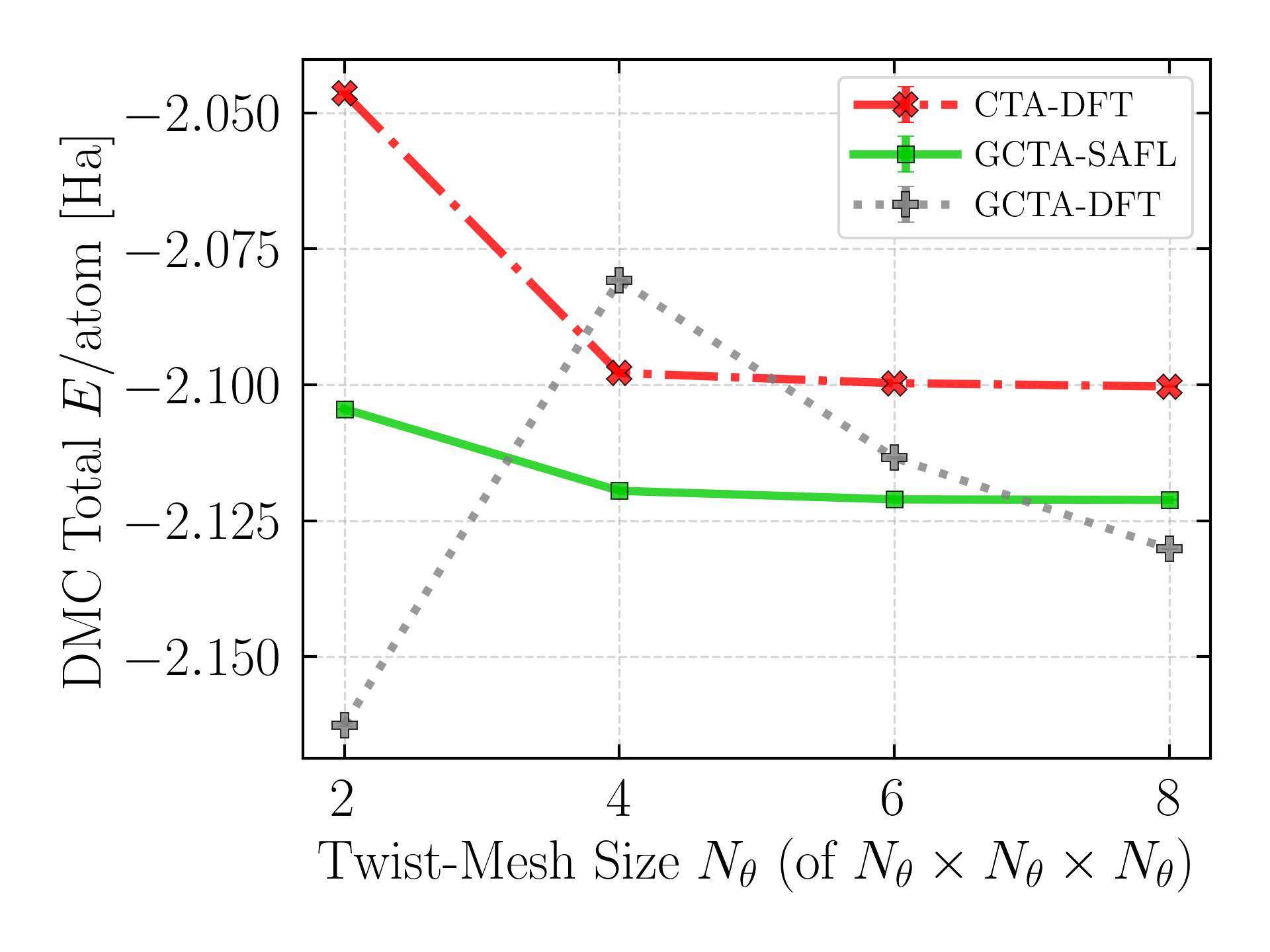}
\caption{$E$/atom}
\label{fig:Al_NM_T1_k_dmc-tot}
\end{subfigure}%
\begin{subfigure}{0.33\textwidth}
\includegraphics[width=\textwidth]{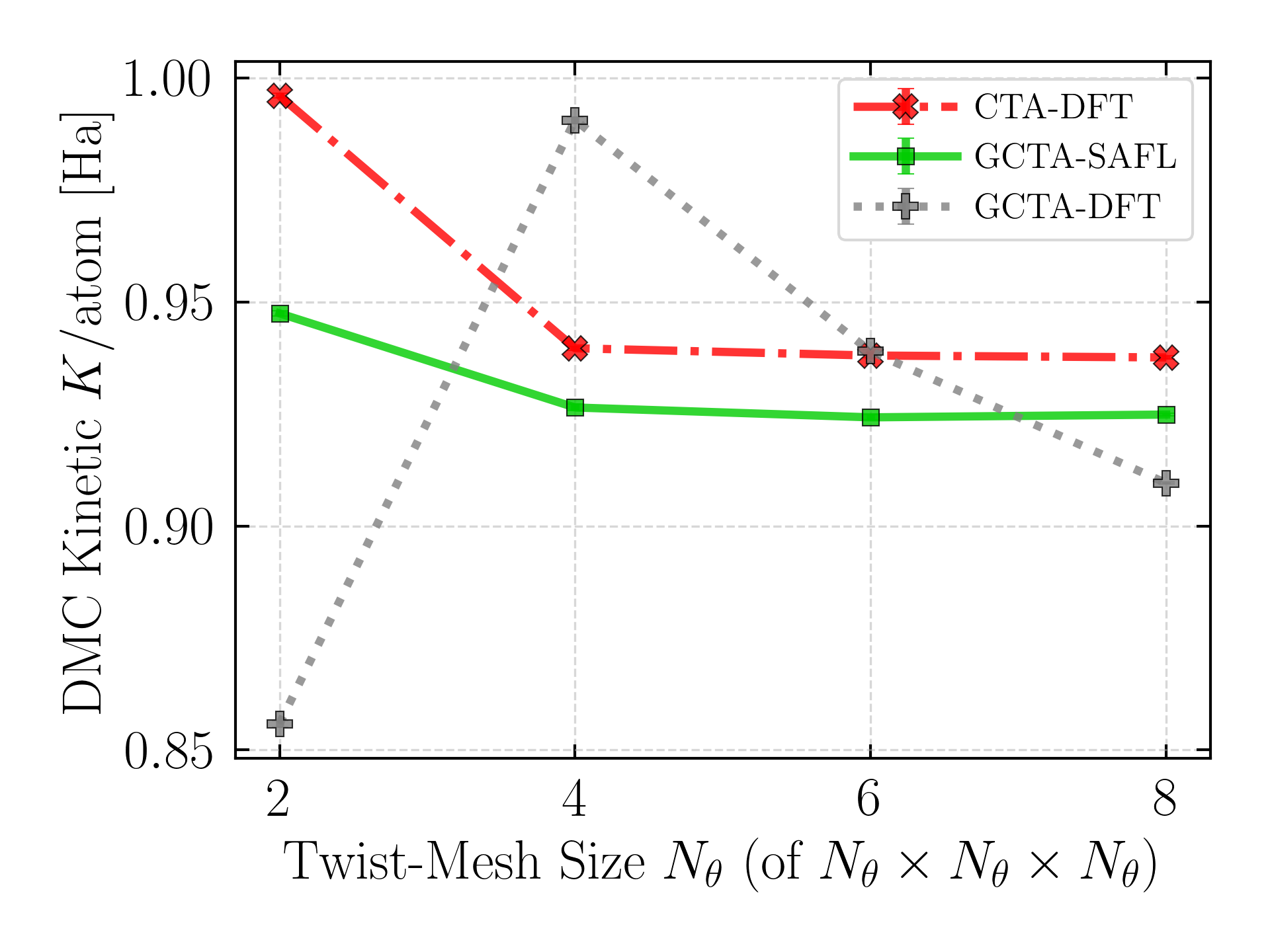}
\caption{$K$/atom}
\label{fig:Al_NM_T1_k_dmc-kin}
\end{subfigure}%
\begin{subfigure}{0.33\textwidth}
\includegraphics[width=\textwidth]{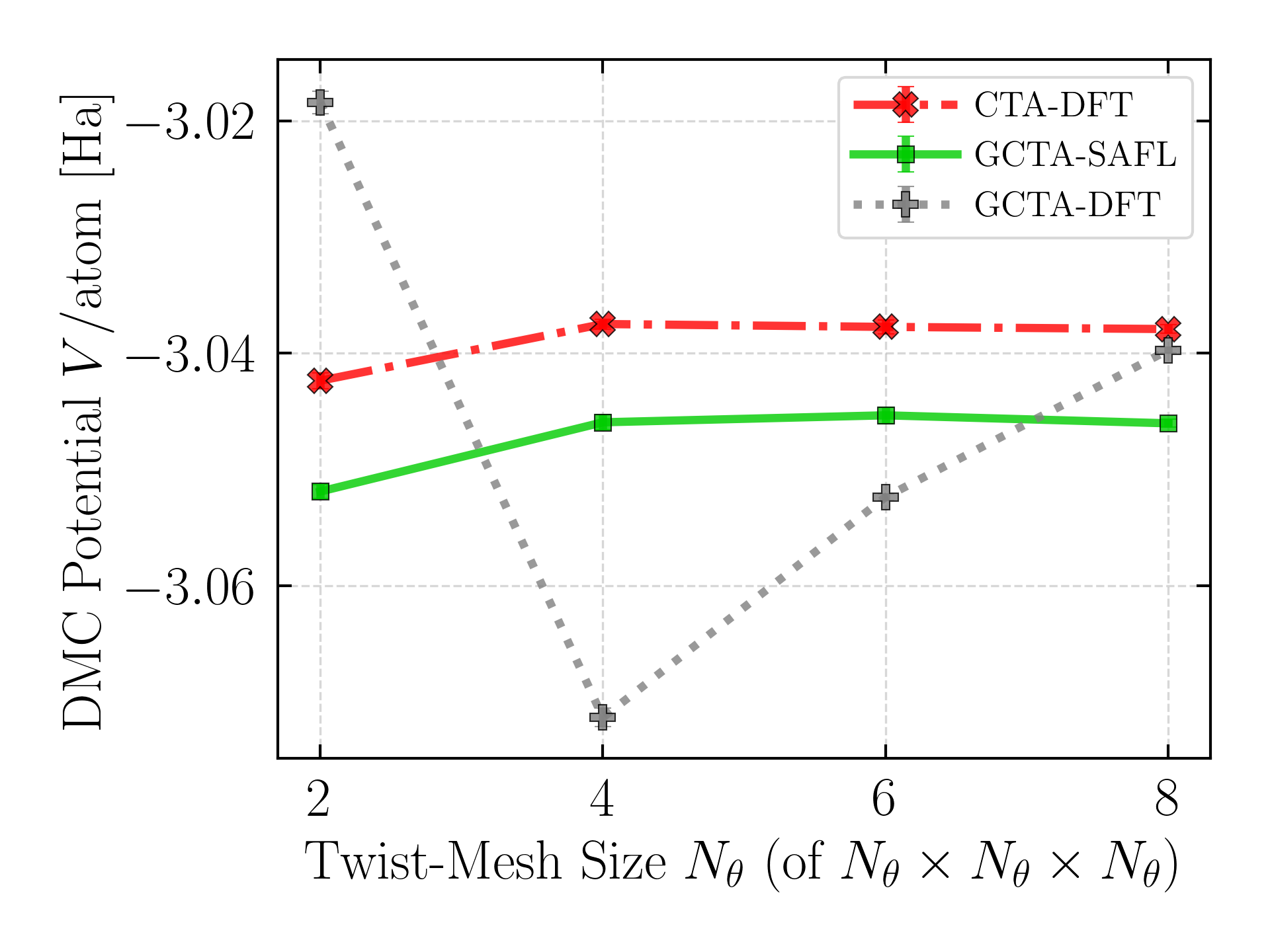}
\caption{$V$/atom}
\label{fig:Al_NM_T1_k_dmc-pot}
\end{subfigure}
\begin{subfigure}{0.33\textwidth}
\includegraphics[width=\textwidth]{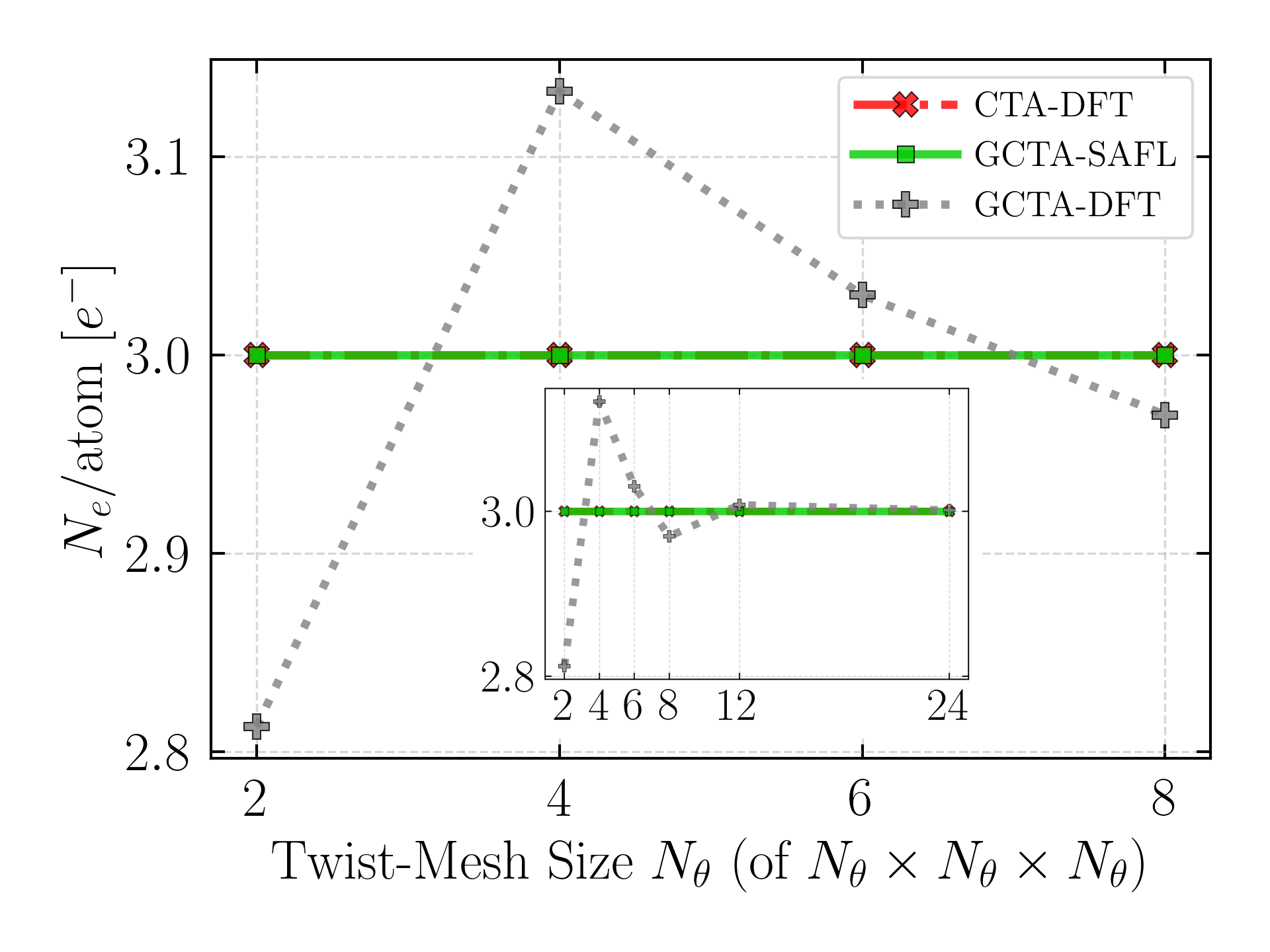}
\caption{$N_e$/atom}
\label{fig:Al_NM_T1_k_occupation}
\end{subfigure}%
\begin{subfigure}{0.33\textwidth}
\includegraphics[width=\textwidth]{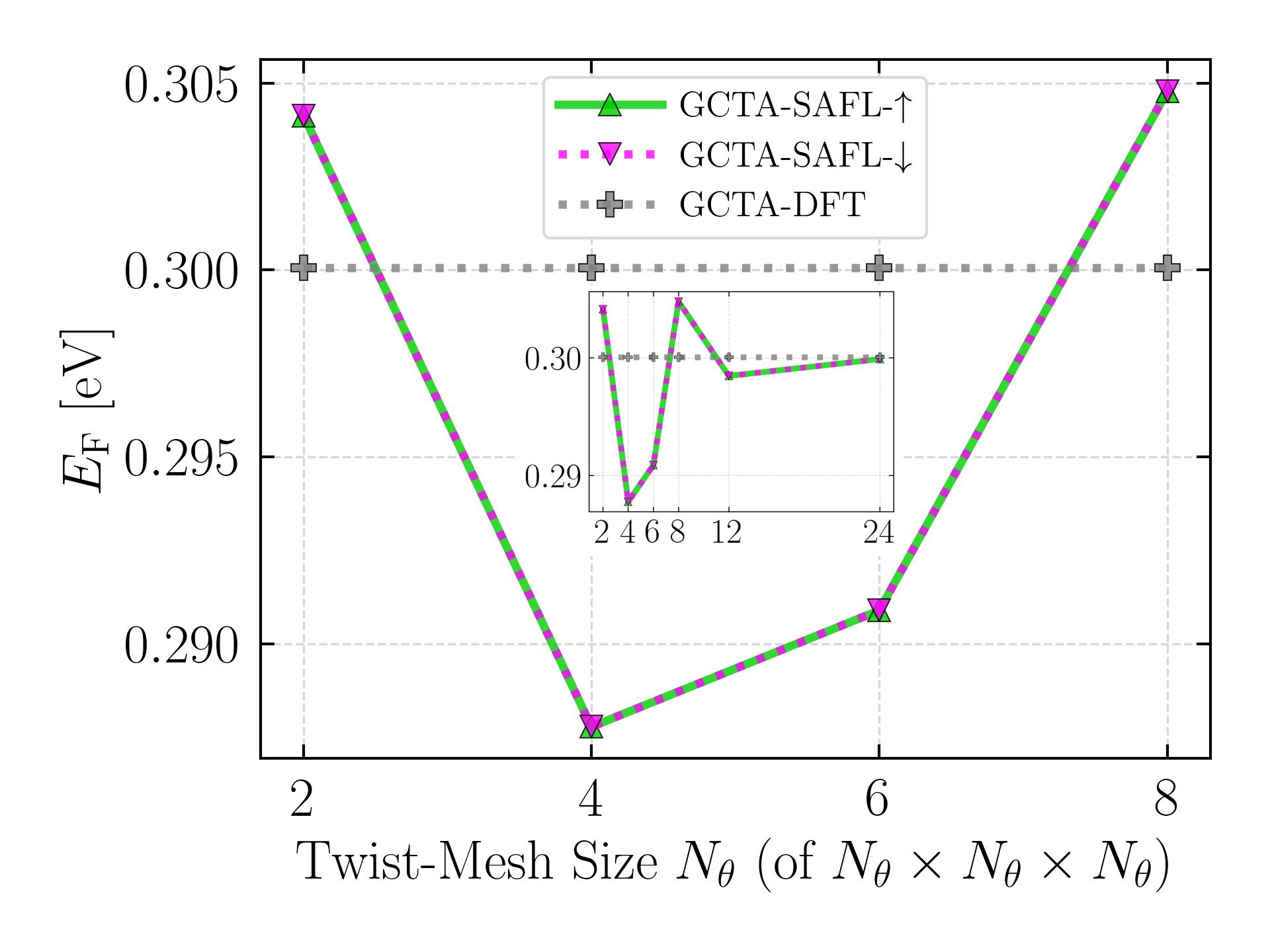}
\caption{$E_\mathrm{F}$}
\label{fig:Al_NM_T1_k_efermi}
\end{subfigure}%
\begin{subfigure}{0.33\textwidth}
\includegraphics[width=\textwidth]{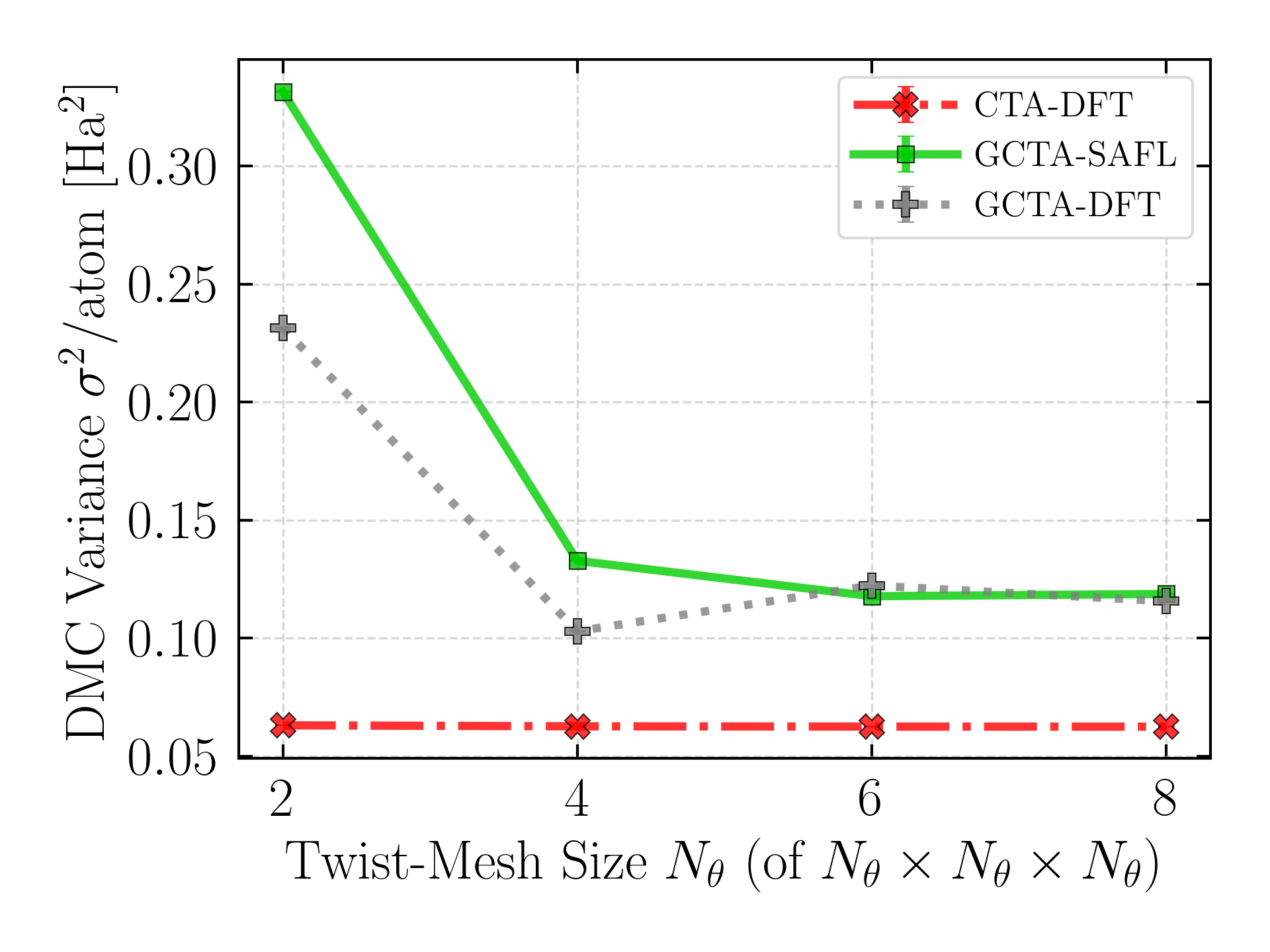}
\caption{$\sigma^2$/atom}
\label{fig:Al_NM_T1_k_dmc-var}
\end{subfigure}%
\caption{
Convergence of various twist-averaged quantities of fcc Al (4 Al atoms) with respect to the employed QMC twist-mesh.
(a-c) DMC energy components,
(d) number of electrons in QMC,
(e) Fermi levels set in QMC, and
(f) DMC total energy variance.
The twist-mesh is uniform in the reciprocal space and includes the high-symmetry points [$\Gamma$, X, M, R].
QMC errors are one standard deviation and smaller than the data symbol sizes.
}
\label{fig:Al_NM_k_mesh_conv}
\end{figure*}

Figure \ref{fig:Al_NM_k_mesh_conv} shows the convergence of various quantities in the 4-atom Al cell as the twist-mesh is increased.
As Al is nonmagnetic, we mainly focus on the convergence of energetics and only show GCTA-SAFL, GCTA-DFT, and CTA-DFT occupation results (GCTA-SAFL vs. GCTA-AFL and CTA-DFT vs. CTA-INS are equivalent).
First, let us consider the DMC energies. 
Figures \ref{fig:Al_NM_T1_k_dmc-tot}, \ref{fig:Al_NM_T1_k_dmc-kin}, \ref{fig:Al_NM_T1_k_dmc-pot} shows the DMC total, kinetic, and potential energies, respectively.
GCTA-SAFL and CTA-DFT energies quickly converge at $N_{\theta} = 4$.
However, they converge to different values: the total energy of GCTA-SAFL is lower than that of CTA-DFT by about $21$ mHa/atom for large $N_{\theta}$.
The lower GCTA-SAFL total energy is not surprising as the occupations are closer to that of real metal.
The total energies of these two schemes can be directly compared since the twist-averaged system is charge-neutral in both cases ($Q_\mathrm{TA} = 0$).
In GCTA-SAFL (and GCTA-AFL), this leads to an exact cancellation of bias due to background charges.

GCTA-DFT shows wildly fluctuating energies.
This is due to the fluctuating number of electrons $N_e$ as shown in Figure \ref{fig:Al_NM_T1_k_occupation}.
For large values of $N_\theta$, $N_e$ approaches the correct number of electrons up to a small discrepancy $\epsilon = 0.001~e^{-}$/atom (Figure \ref{fig:Al_NM_T1_k_occupation}, inset), which arises due to fractional occupations in DFT.
In GCTA-SAFL, these large $N_e$ fluctuations are controlled by a fluctuating $E_\mathrm{F}$ as shown in Figure \ref{fig:Al_NM_T1_k_efermi}.
In this case, up and down SAFL Fermi levels are the same since the spin channels are equivalent.
Again, for large values of $N_{\theta}$, GCTA-SAFL Fermi levels approach the DFT value up to a small discrepancy (inset of Figure \ref{fig:Al_NM_T1_k_efermi}).

Finally, it is worth commenting on the cost of different occupation schemes.
Figure \ref{fig:Al_NM_T1_k_dmc-var} shows the twist-averaged DMC total energy variances. 
CTA-DFT shows a flat and relatively low variance as $N_\theta$ is increased.
On the other hand, both GCTA methods start with very large values for small $N_\theta$ and converge to about double the variance of CTA-DFT for large $N_\theta$.
The increase in variance is due to charge variation per twist in GCTA ($q(\theta) \neq 0$)), which is not present in CTA ($q(\theta) = 0$)).
Therefore, GCTA methods require more samples to reach the same error bar as those of the CTA methods.
The moderate increase in cost is well justified, as shown in the TDL extrapolations in Section \ref{sec:supercell}.

\subsubsection{Fe}

\begin{figure*}[!htbp]
\centering
\begin{subfigure}{0.33\textwidth}
\includegraphics[width=\textwidth]{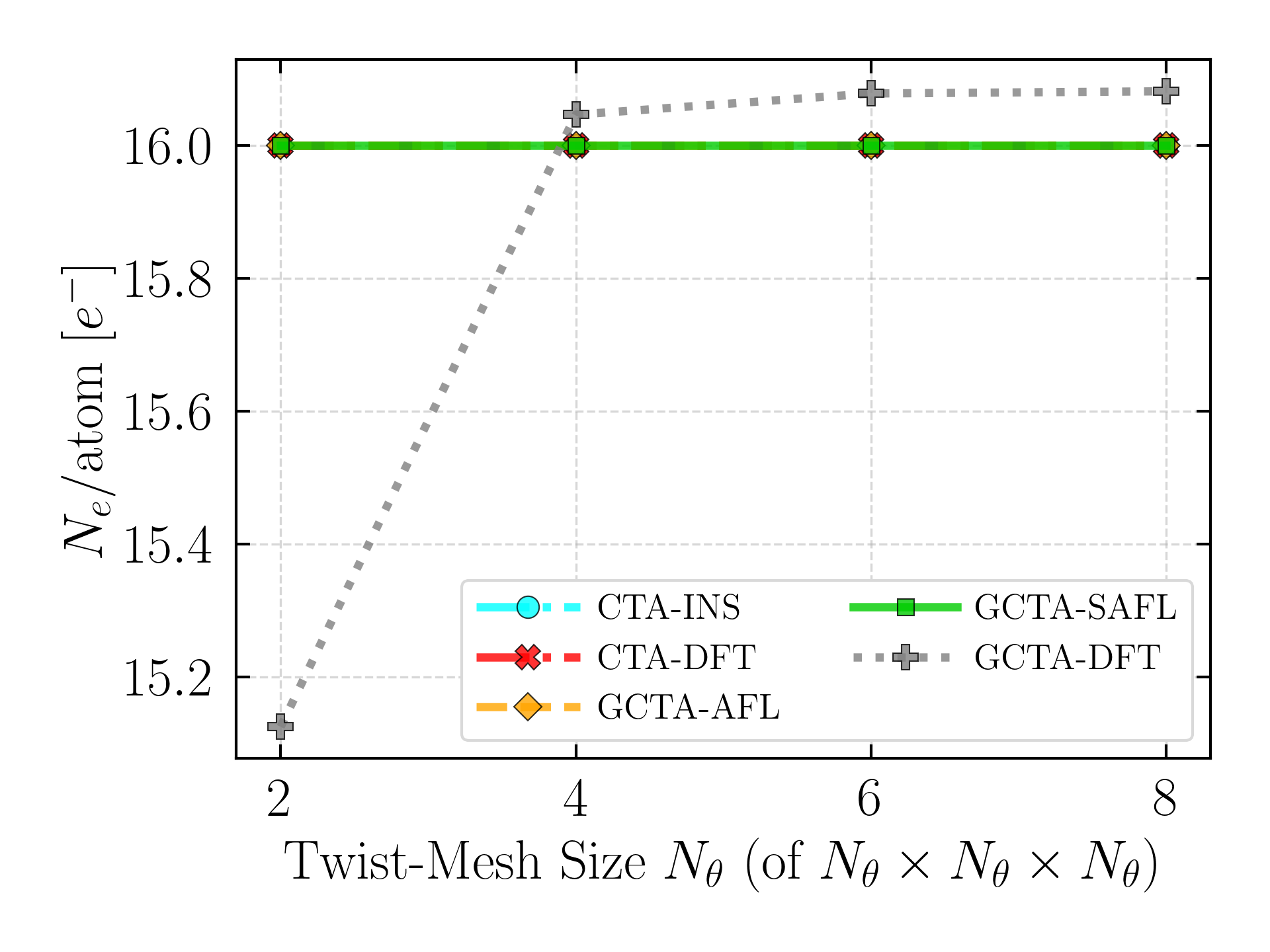}
\caption{$N_e$/atom}
\label{fig:Fe_FM_T1_k_occupation}
\end{subfigure}%
\begin{subfigure}{0.33\textwidth}
\includegraphics[width=\textwidth]{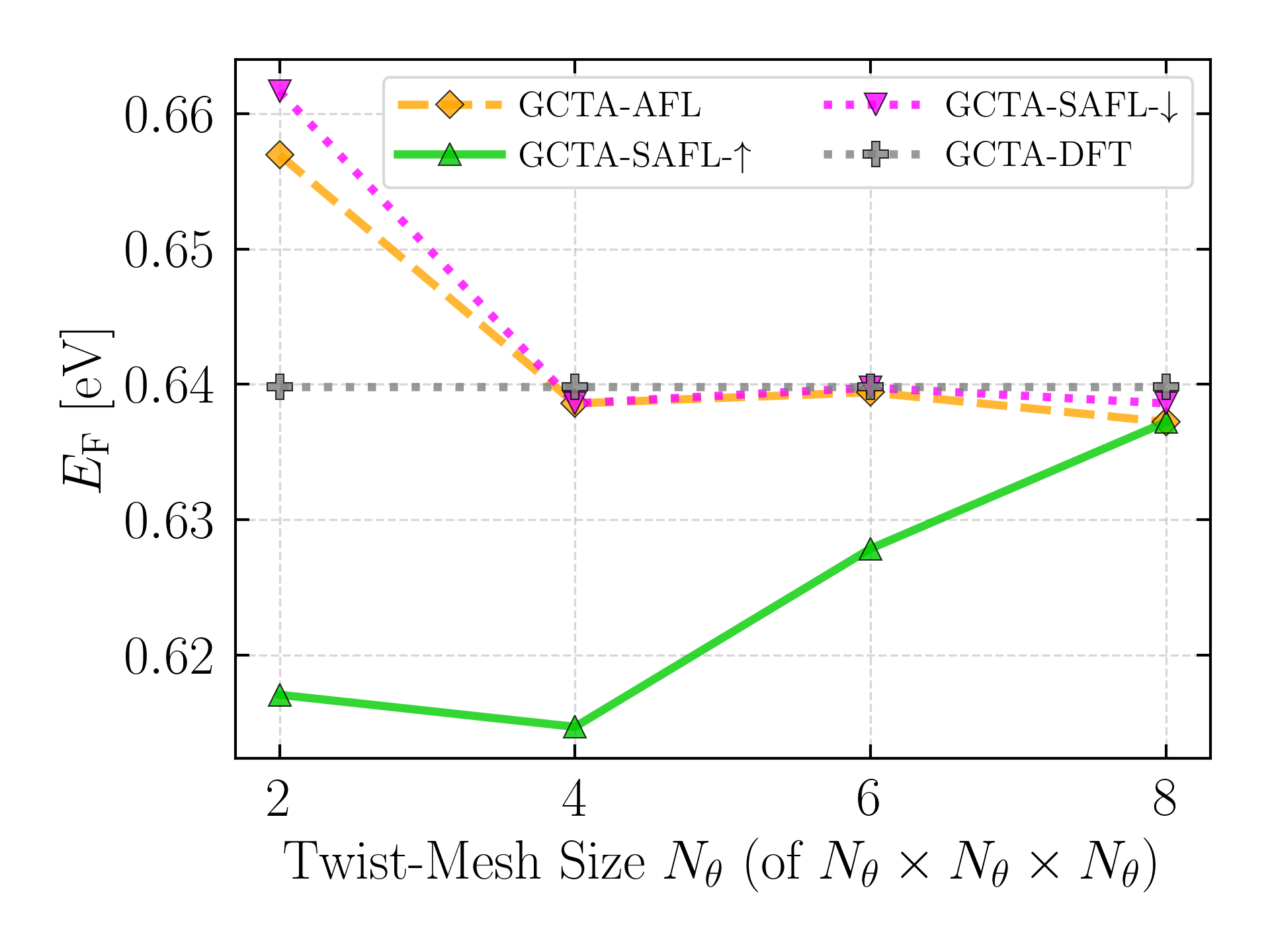}
\caption{$E_\mathrm{F}$}
\label{fig:Fe_FM_T1_k_efermi}
\end{subfigure}%
\begin{subfigure}{0.33\textwidth}
\includegraphics[width=\textwidth]{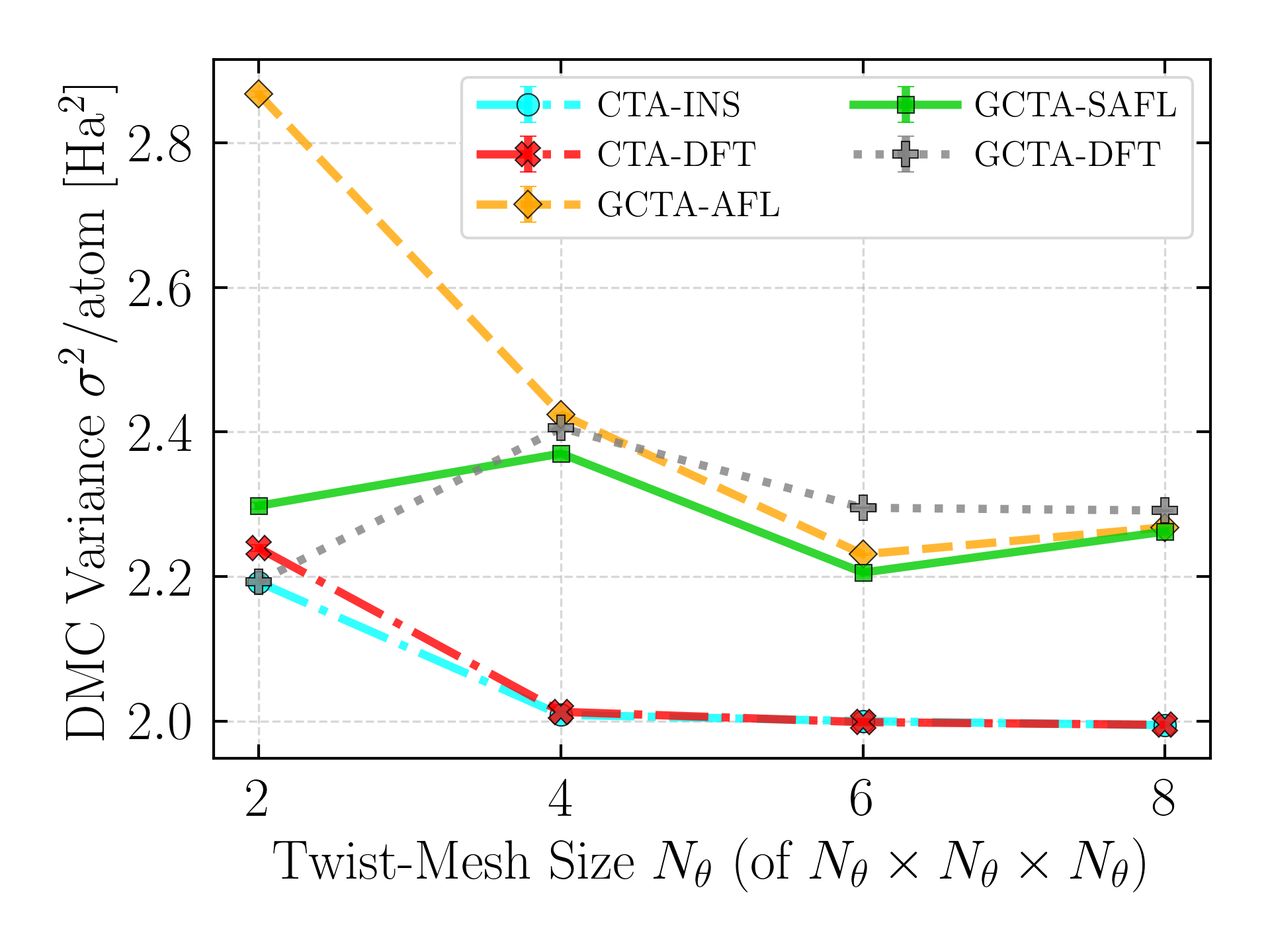}
\caption{$\sigma^2$/atom}
\label{fig:Fe_FM_T1_k_dmc-var}
\end{subfigure}
\begin{subfigure}{0.33\textwidth}
\includegraphics[width=\textwidth]{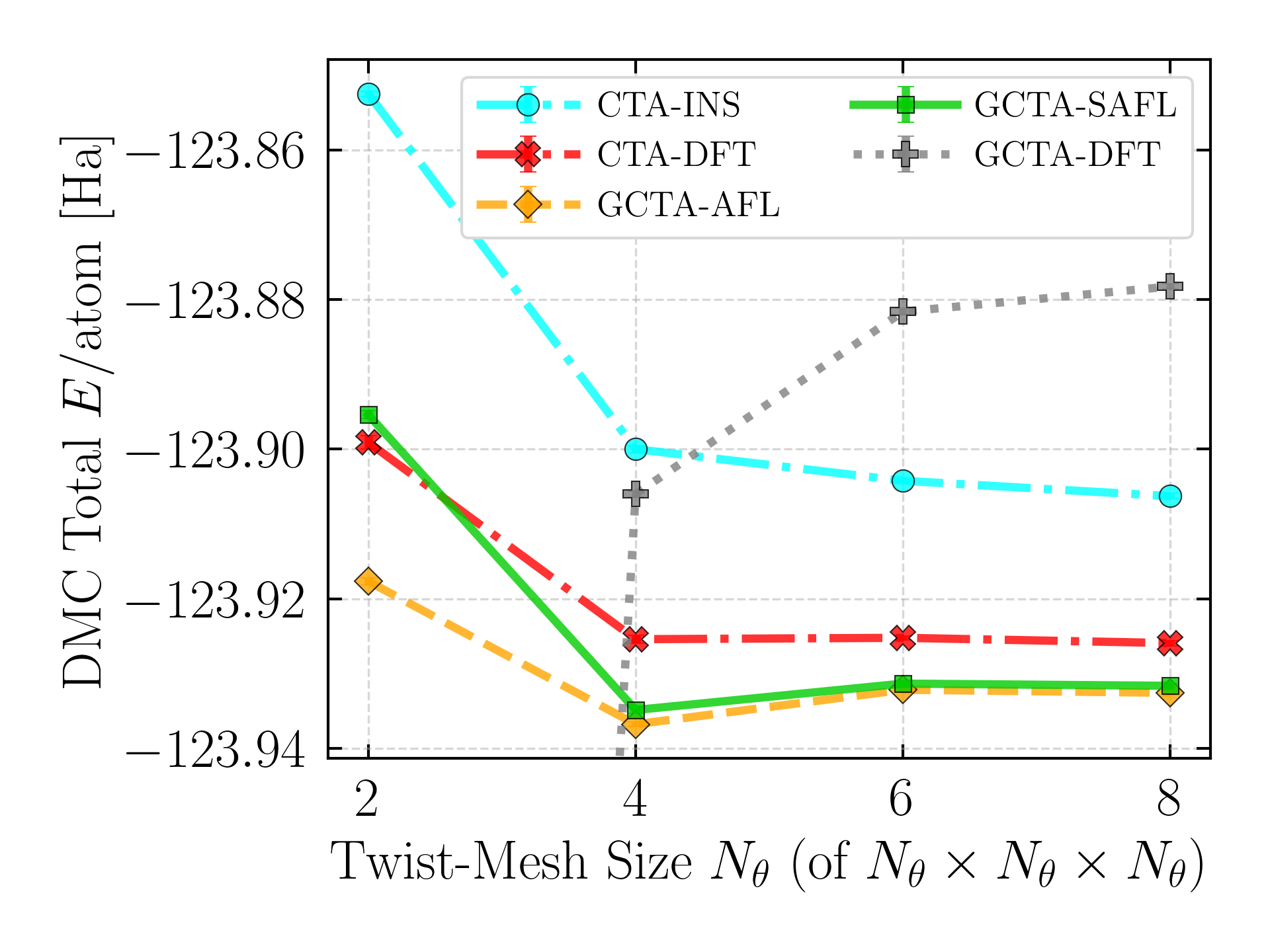}
\caption{$E$/atom}
\label{fig:Fe_FM_T1_k_dmc-tot}
\end{subfigure}%
\begin{subfigure}{0.33\textwidth}
\includegraphics[width=\textwidth]{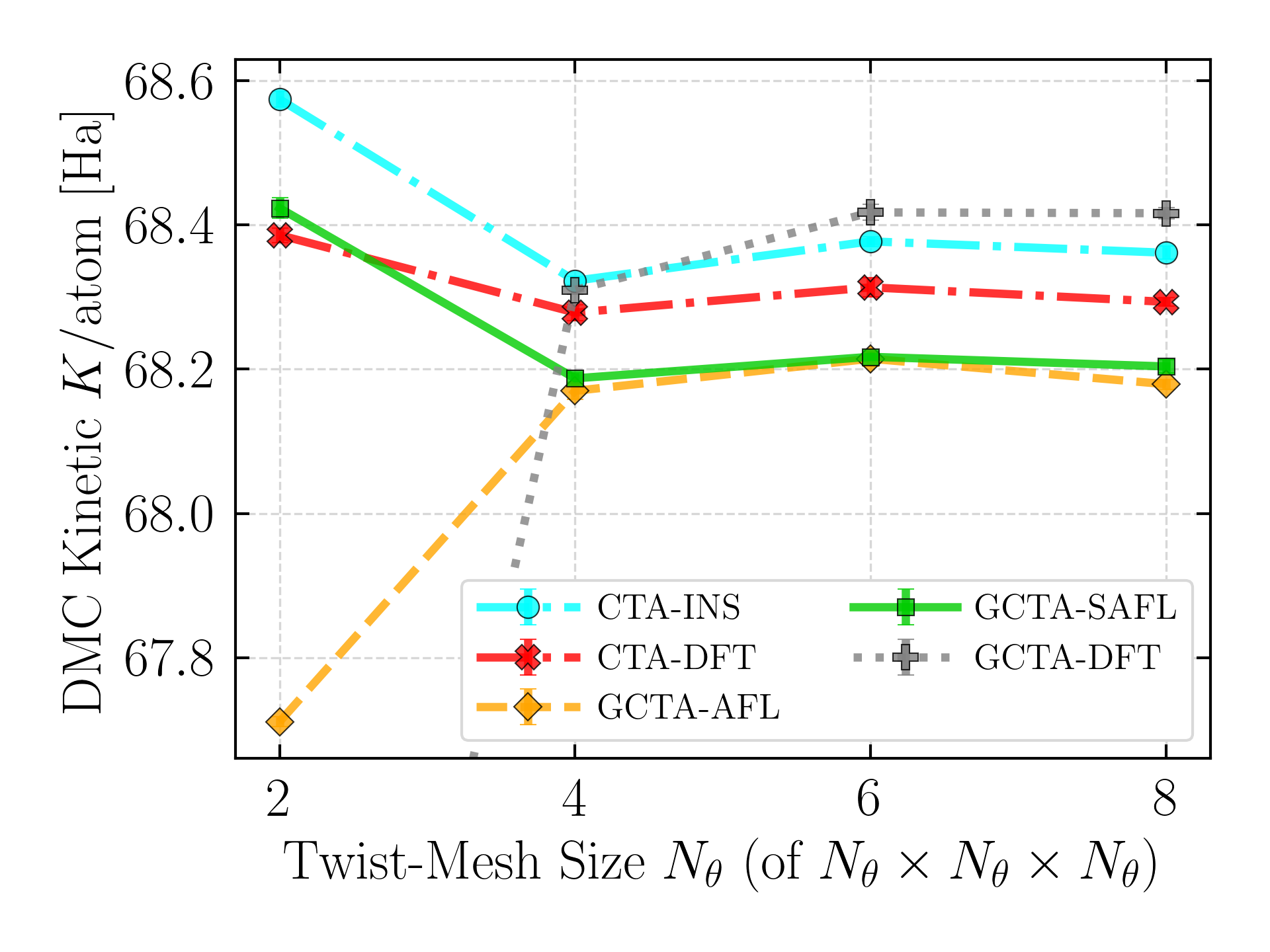}
\caption{$K$/atom}
\label{fig:Fe_FM_T1_k_dmc-kin}
\end{subfigure}%
\begin{subfigure}{0.33\textwidth}
\includegraphics[width=\textwidth]{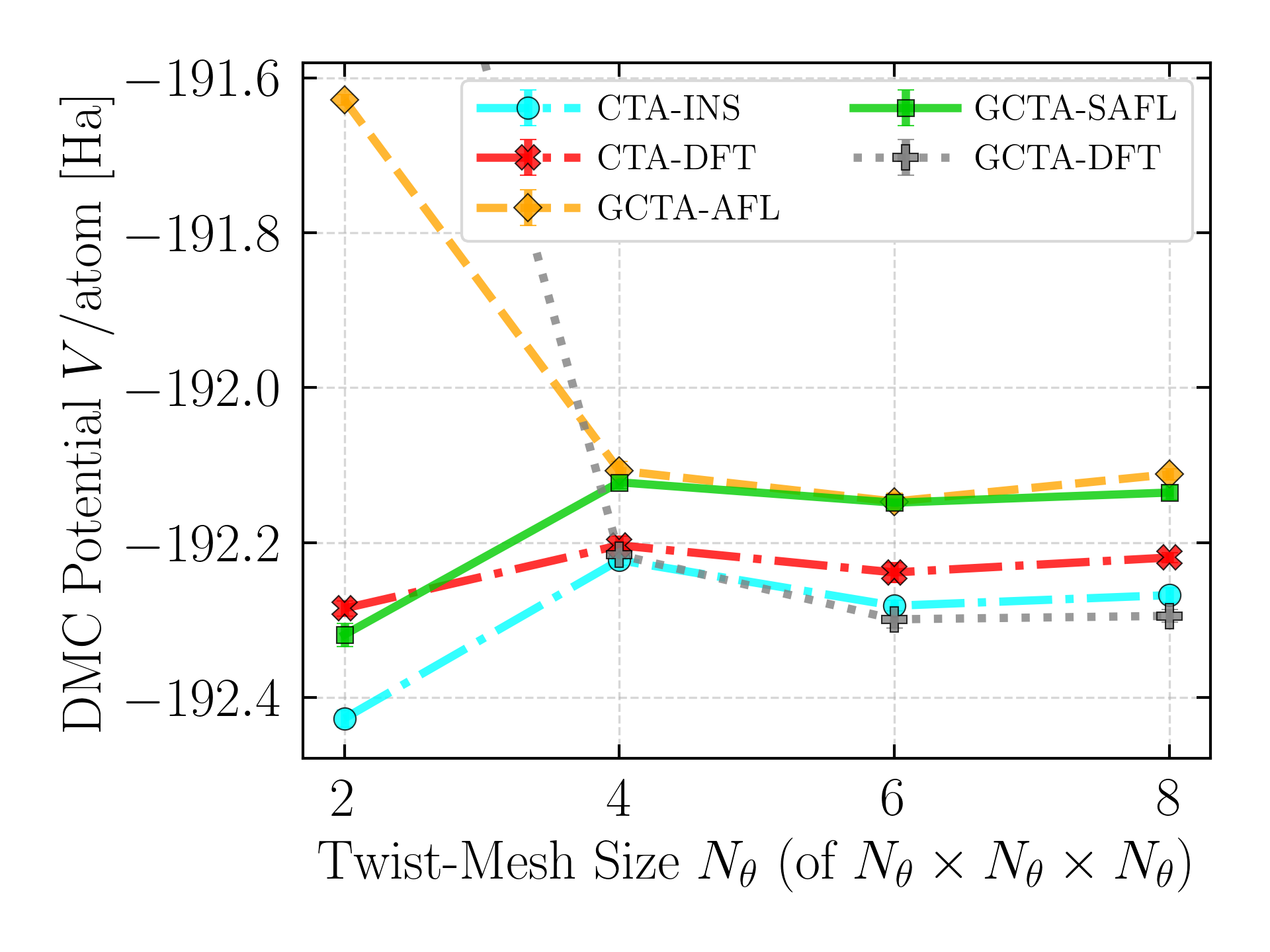}
\caption{$V$/atom}
\label{fig:Fe_FM_T1_k_dmc-pot}
\end{subfigure}
\begin{subfigure}{0.33\textwidth}
\includegraphics[width=\textwidth]{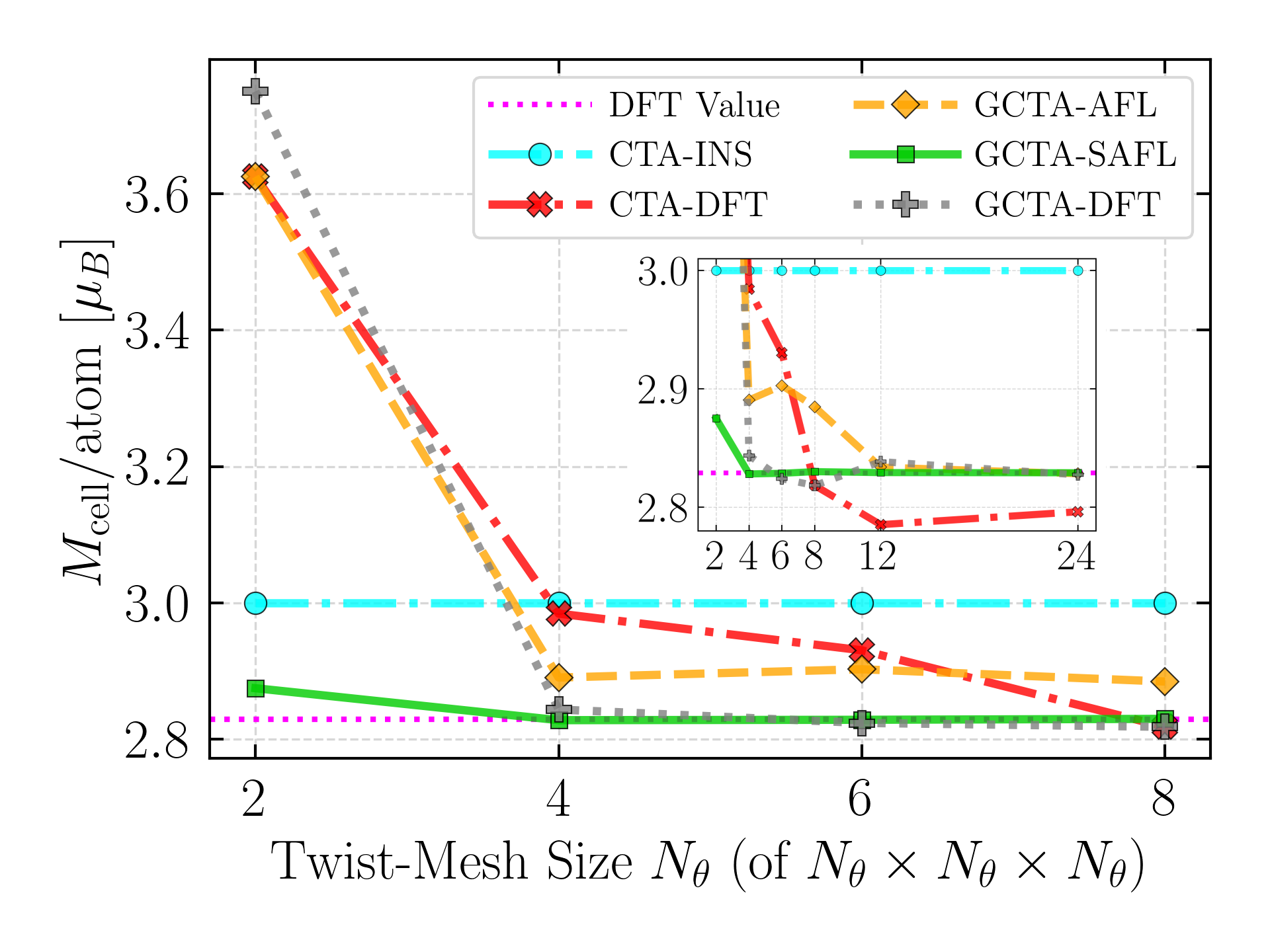}
\caption{$M_\mathrm{cell}$/atom}
\label{fig:Fe_FM_T1_k_magnetization}
\end{subfigure}%
\begin{subfigure}{0.33\textwidth}
\includegraphics[width=\textwidth]{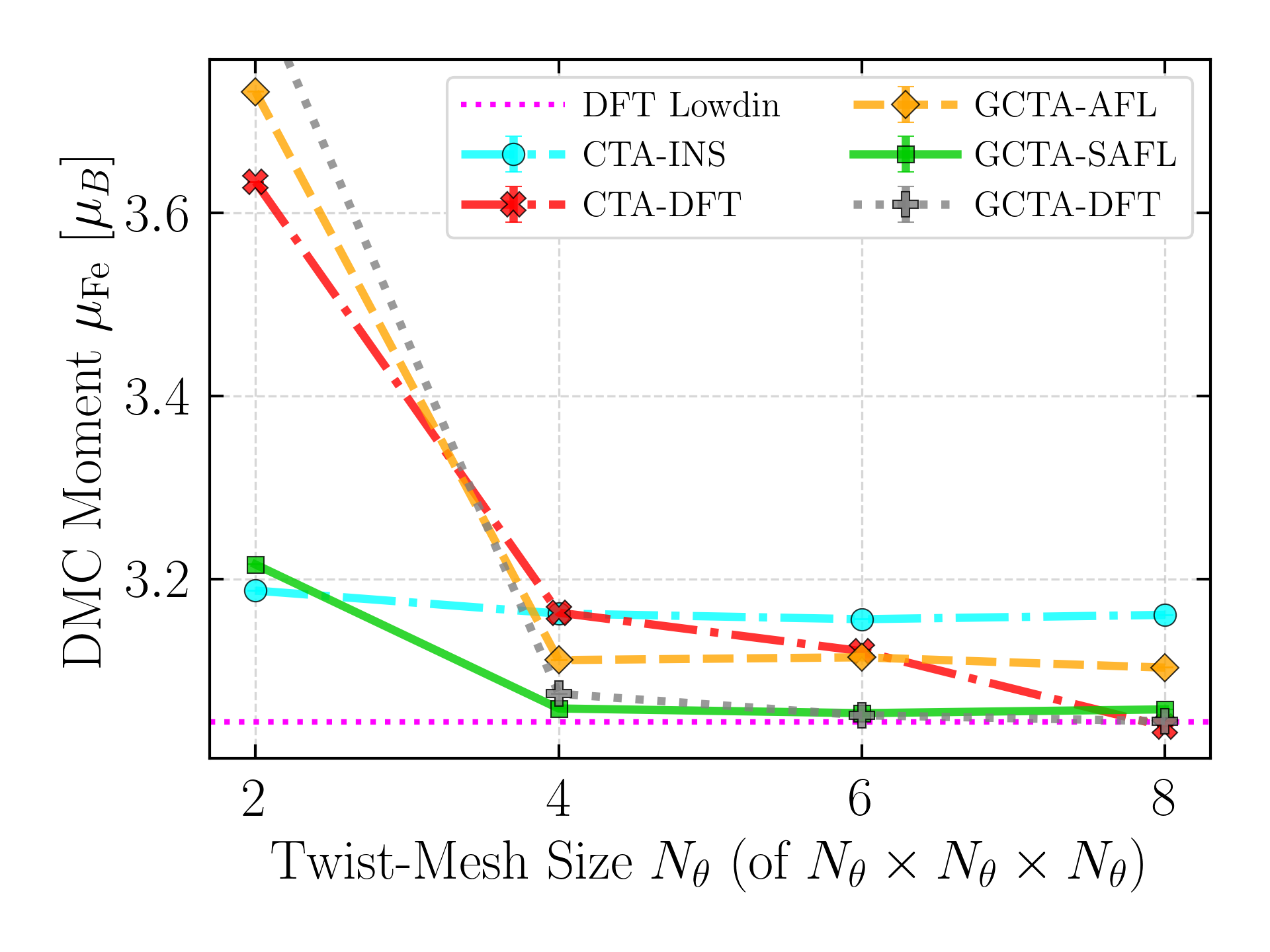}
\caption{$\mu_\mathrm{Fe}$}
\label{fig:Fe_FM_T1_k_dmc-mom}%
\end{subfigure}
\begin{subfigure}{0.33\textwidth}
\includegraphics[width=\textwidth]{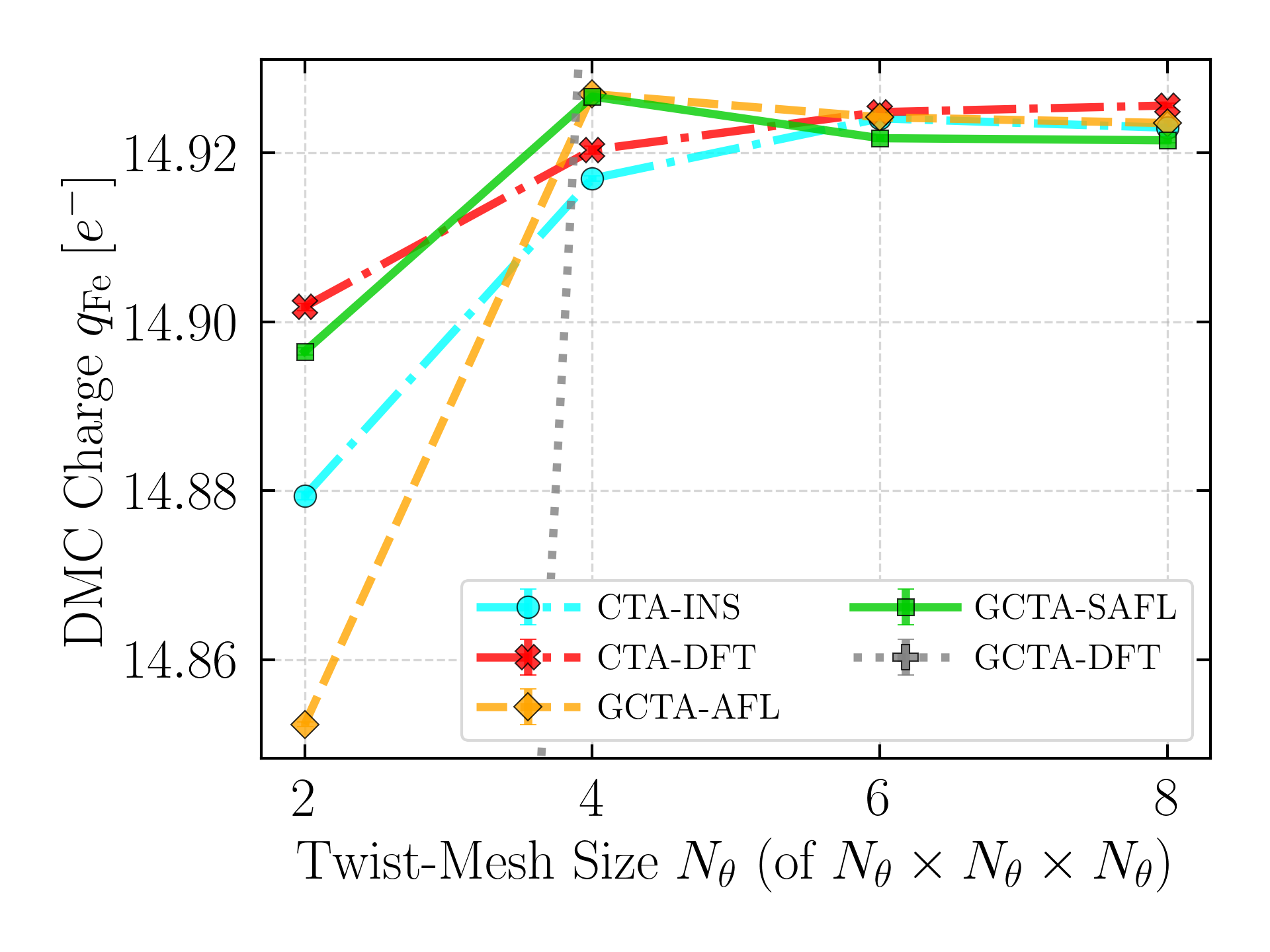}
\caption{$q_\mathrm{Fe}$}
\label{fig:Fe_FM_T1_k_dmc-chg}
\end{subfigure}
\caption{
Convergence of various twist-averaged quantities of bcc Fe (2 Fe atoms) with respect to the employed QMC twist-mesh.
(a) Number of electrons in QMC,
(b) Fermi levels set in QMC,
(c) DMC total energy variance,
(d-f) DMC energy components,
(g) cell magnetization,
(h) DMC atomic moment, and
(i) DMC atomic charge.
The twist-mesh is uniform in the reciprocal space and includes the high-symmetry points [$\Gamma$, X, M, R].
QMC errors are one standard deviation and smaller than the data symbol sizes.
}
\label{fig:Fe_FM_k_mesh_conv}
\end{figure*}

Next, let us consider the convergence of quantities in $\alpha$-Fe, Figure \ref{fig:Fe_FM_k_mesh_conv}.
In this case, the spin channels are not equivalent, and we include all occupation schemes.
Overall, we observe a behavior similar to that of the Al case.
All TA schemes result in $Q_\mathrm{TA} = 0$ except for GCTA-DFT as shown in Figure \ref{fig:Fe_FM_T1_k_occupation}.
GCTA-AFL and GCTA-SAFL achieve charge-neutrality by allowing a fluctuating $E_\mathrm{F}$, and they approach the GCTA-DFT $E_\mathrm{F}$ up to a small difference for large $N_\theta$; see Figure \ref{fig:Fe_FM_T1_k_efermi}.
Note that GCTA-SAFL up and down Fermi levels have different values here to target the reference magnetization.
As previously reported, all GCTA variances converge to larger values than CTA schemes, albeit the increase, in this case, is not as dramatic, Figure \ref{fig:Fe_FM_T1_k_dmc-var}.

Figures \ref{fig:Fe_FM_T1_k_dmc-tot}, \ref{fig:Fe_FM_T1_k_dmc-kin}, \ref{fig:Fe_FM_T1_k_dmc-pot} shows the DMC total, kinetic, and potential energies, respectively.
In all cases, the energies are converged for $N_\theta \geq 6$.
Again, GCTA-SAFL and GCTA-AFL obtained the lowest total energies among the TA methods.
GCTA-SAFL and GCTA-AFL total energies differ by less than 1~mHa/atom, which is most likely due to slightly different magnetizations obtained in the QMC.

Figure \ref{fig:Fe_FM_T1_k_magnetization} shows the cell magnetizations set by various occupation schemes.
The dotted horizontal line is the value obtained from the dense $k$-mesh DFT calculation, which is used here as the target value.
Related to cell magnetization, Figure \ref{fig:Fe_FM_T1_k_dmc-mom} shows the Fe atomic moments calculated in DMC (see Section \ref{sec:computation} and the SI for details).
Again, the dotted horizontal line is the DFT moment using L\"owdin occupations.
These plots show that GCTA-SAFL has the best convergence in the magnetism of the system, both in cell magnetization and atomic moments.
Surprisingly, GCTA-DFT magnetizations are also accurate for $N_\theta \geq 4$, although the total energies are quite off.
Also, CTA-DFT magnetization for large values of $N_\theta$ is reasonably close to the target value despite not explicitly including it in the recipe.  
In GCTA-AFL with moderate values of $N_\theta = [4,6,8]$, there is a deviation of magnetism by about 0.05~$\mu_\mathrm{B}$, due to shifting both spin channels uniformly.
However, for a large $N_\theta$, GCTA-AFL agrees well with GCTA-SAFL (inset of Figure \ref{fig:Fe_FM_T1_k_magnetization}).
CTA-INS cell magnetization is fixed at integer value $M_\mathrm{cell} = 3~\mu_\mathrm{B}$ since every twist must have the same polarization.
Finally, Figure \ref{fig:Fe_FM_T1_k_dmc-chg} shows the atomic charge contained within a certain radius.
All occupation schemes except for GCTA-DFT converge to a similar value of $14.92~e^{-}$.

Considering the energetics in Al and $\alpha$-Fe, as well as the magnetism in $\alpha$-Fe, it is fair to conclude that GCTA-SAFL shows the most robust convergence in DMC observables when $N_\theta$ is increased within the primitive cell.
GCTA-AFL is also quite competitive, although the resulting magnetic phase converges slowly.
However, in the case of low Hubbard $U$ values and thus small atomic moments, the distinction between GCTA-SAFL and GCTA-AFL diminishes, becoming equivalent at zero magnetic moments.
For example, our preliminary runs with LDA for $\alpha$-Fe show that GCTA-AFL energies and moments converge almost as fast as those for GCTA-SAFL.
Nevertheless, having control over a wide range of $U$ values is an advantage for GCTA-SAFL when studying the impact of $U$ in the wave function nodal surface.

\subsection{Supercell Size Convergence}
\label{sec:supercell}

Within independent-electron theories such as 
KS-DFT, the TDL can be reached by increasing the $k$-mesh in the primitive cell since the Hamiltonian is invariant under the translation of a single electron \cite{kratzer_basics_2019}.
In real-space QMC, a many-body theory with electron-electron interactions, we use the ``supercell Hamiltonian", which imposes \textit{artificial} translational symmetry of any electron by $\mathbf{R}_\mathrm{s}$ \cite{rajagopal_quantum_1994, rajagopal_variational_1995, foulkes_quantum_2001}.
This is not a symmetry of a truly infinite Hamiltonian.
Therefore, reaching the TDL via supercell extrapolation is crucial, ideally with the smallest supercells and robust $1/N_\mathrm{atom}$ linear extrapolations ($N_\mathrm{atom}$ is the number of atoms in a supercell).
We have demonstrated that GCTA-SAFL converges to energies lower than those of CTA methods while improving the convergence within the smallest cell ($Z_T = 1$).
Here, we compare it against other TA schemes when the supercell size is increased.
The desired TA scheme should result in the best linear relationship for the energies such that only the smallest cells can be utilized for TDL estimation.
Namely, the slope using the small cells should be consistent with the slope from the larger cells.
The importance of this simple criterion can be gleaned from metals and insulators with $\Gamma$-point sampling.
The ground state slope of the energy in such calculations can change drastically due to shell-filling effects, to the extent that even the sign of the slope can change \cite{dagrada_exact_2016, annaberdiyev_cohesion_2021}.
A wild slope change is problematic and less suitable for statistically robust fitting and extrapolation.
In addition, it is also desirable to have the slope $\textit{values}$ to be close to zero (flat extrapolations) facilitated by corrections on potential and kinetic energies \cite{chiesa_finite-size_2006, fraser_finite-size_1996, williamson_elimination_1997, kent_finite-size_1999}.
However, again, this is not the focus of this work, and we report only the raw quantities here.

To investigate the linearity aspect, we carry out calculations with $Z_T = [1^3, 2^3, 3^3]$ supercells, which is enough to comment on the quality of the extrapolations.
For all TA schemes, we chose to keep the total $k$-points employed for tiling the supercells constant ($Z_T \cdot Z_\theta$ = constant).
Figures \ref{fig:Al_NM_k_mesh_conv} and \ref{fig:Fe_FM_k_mesh_conv} show that energies and magnetic properties are converged for GCTA-SAFL at $N_\theta = 6$; thus, we use $Z_T \cdot Z_\theta = 6^3$ in all cases.
To quantify the linearity of each TA scheme, we share the standard deviation of slope $\sigma_\mathrm{S}$ and the commonly used goodness-of-fit metric $R^2$ (coefficient of determination).

\subsubsection{Al}

\begin{figure*}[!htbp]
\centering
\begin{subfigure}{0.33\textwidth}
\includegraphics[width=\textwidth]{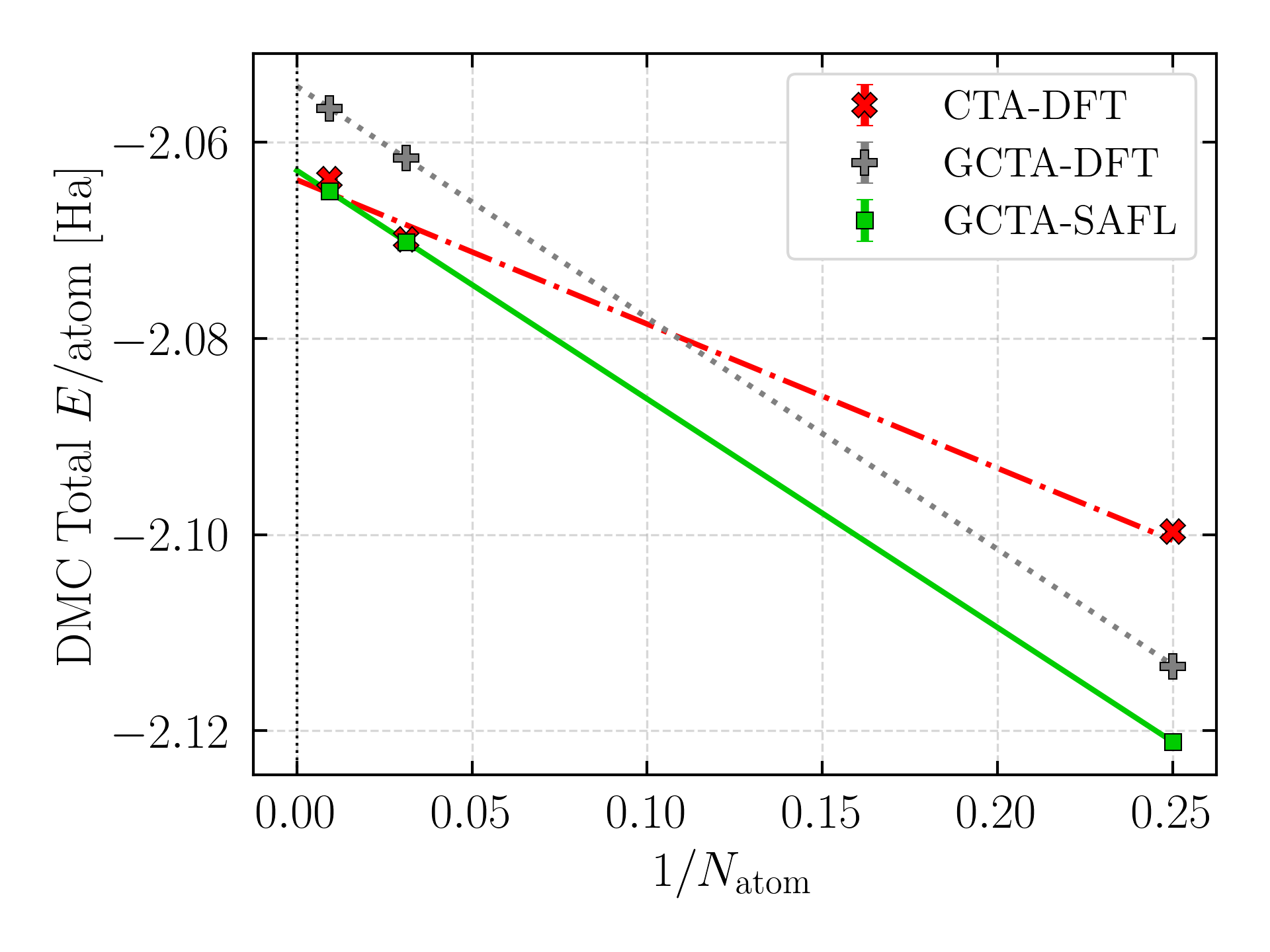}
\caption{$E$/atom}
\label{fig:Al_NM_TDL_dmc-tot}
\end{subfigure}%
\begin{subfigure}{0.33\textwidth}
\includegraphics[width=\textwidth]{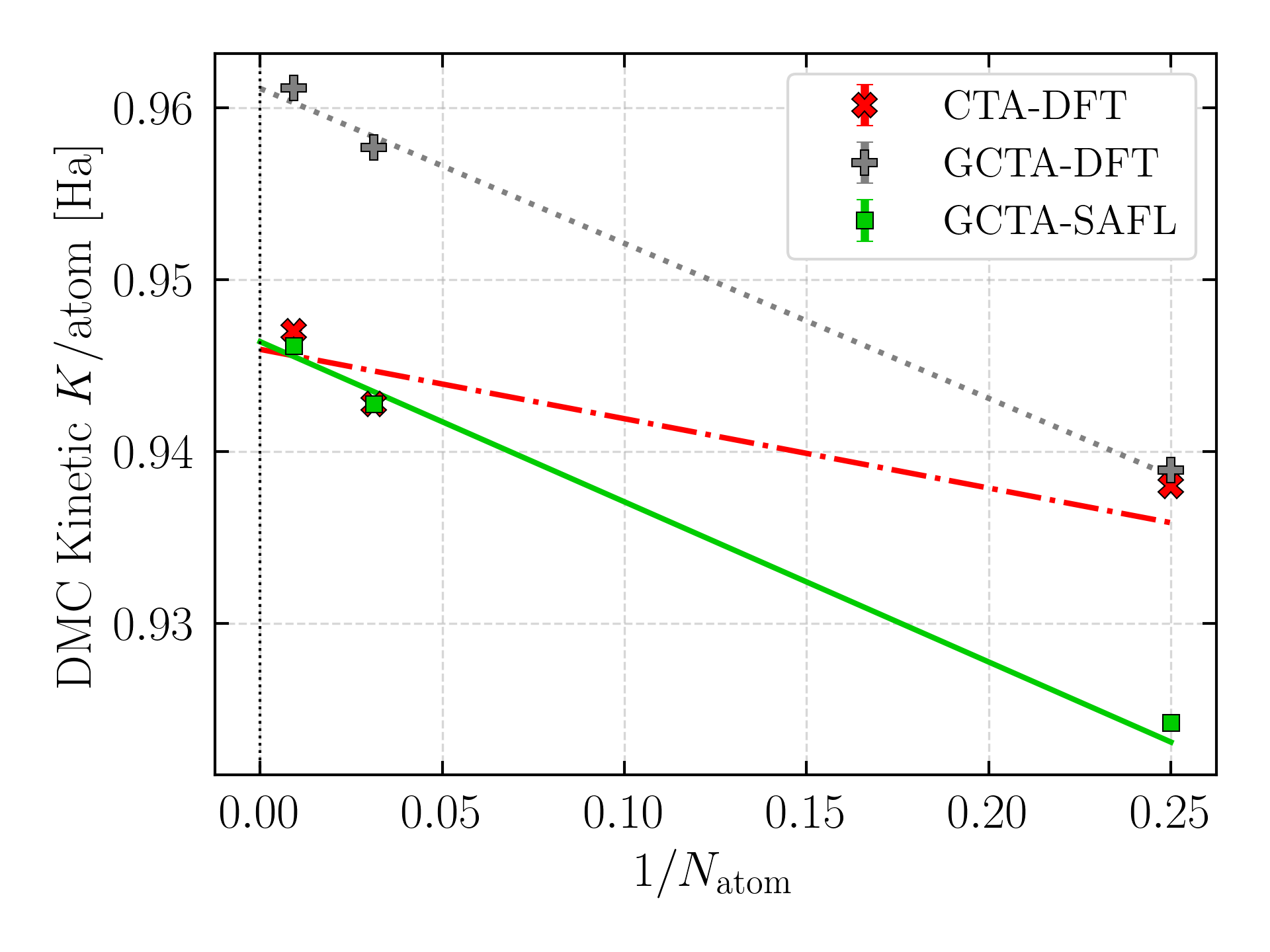}
\caption{$K$/atom}
\label{fig:Al_NM_TDL_dmc-kin}
\end{subfigure}%
\begin{subfigure}{0.33\textwidth}
\includegraphics[width=\textwidth]{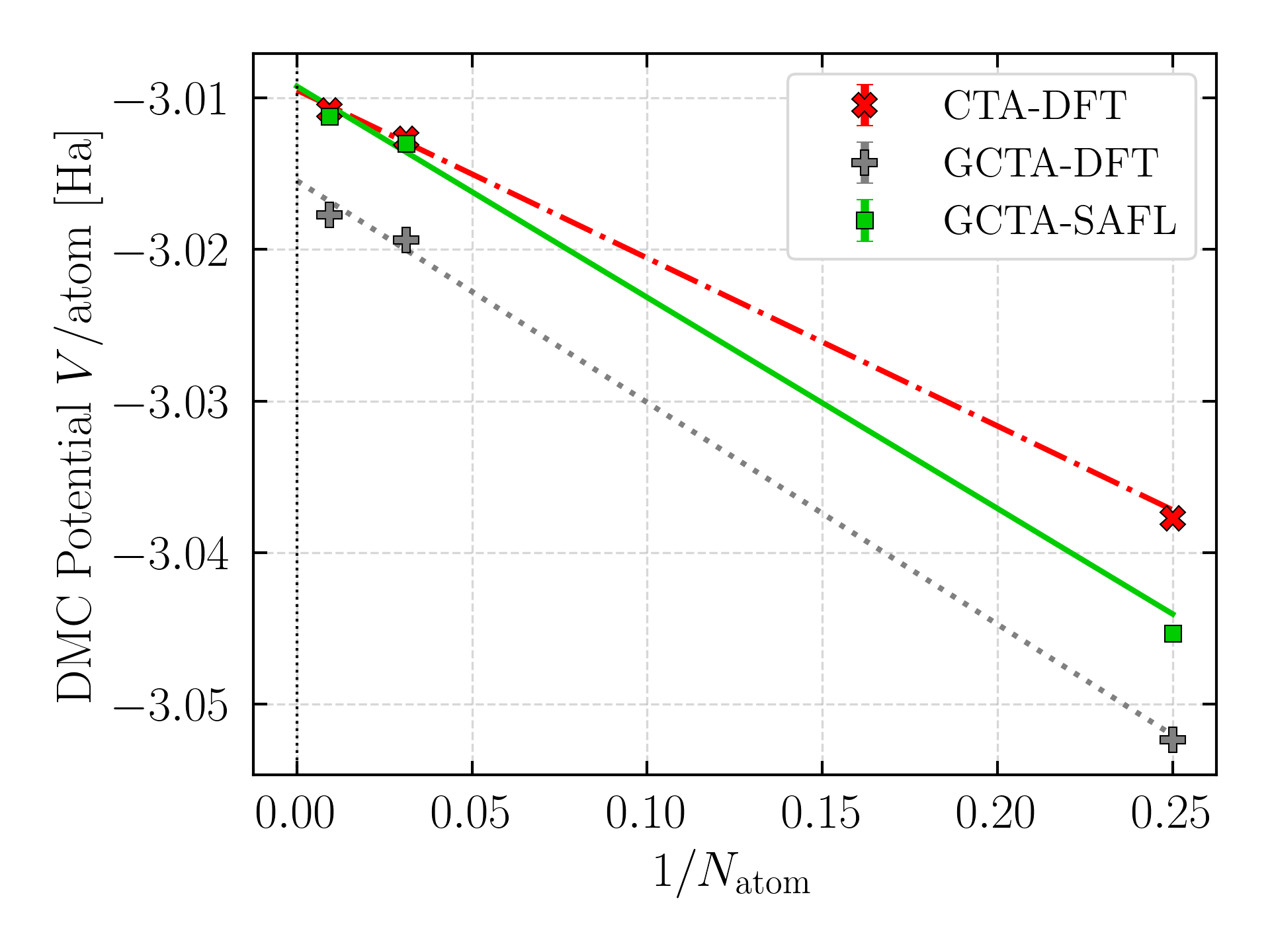}
\caption{$V$/atom}
\label{fig:Al_NM_TDL_dmc-pot}
\end{subfigure}
\caption{
TDL convergence of twist-averaged DMC energies of fcc Al as the supercell is expanded.
(a) DMC total energy,
(b) DMC kinetic energy, and
(c) DMC potential energy.
$Z_T\cdot Z_\theta = 6^3$ is kept constant as the supercell is increased. 
QMC errors are one standard deviation and smaller than the data symbol sizes.
}
\label{fig:Al_NM_TDL_conv}
\end{figure*}

Figure \ref{fig:Al_NM_TDL_conv} shows the DMC total, kinetic, and potential energies as the supercell size increases in Al.
First, we notice that GCTA-DFT converges to different TDL values due to incorrect $N_e$, which is off by about $0.03~e^{-}$; see Figure \ref{fig:Al_NM_T1_k_occupation} at $N_\theta = 6$.
On the other hand, GCTA-SAFL and CTA-DFT TDL values are consistent (with about a 1~mHa difference).
However, large differences are present in the smallest cell.
Overall, we see that the GCTA-SAFL values are more consistent with linear trends.
The linear extrapolation is near-ideal in total energy and slightly worse in energy components.

Table \ref{tab:Al_TDL_fit} provides the metrics for the quality of linear fits.
Here, the bold text represents the least error values.
The total energy shows the best fit with GCTA-SAFL.
The least kinetic and potential energy fit errors correspond to GCTA-DFT and CTA-DFT, respectively.
Despite the better linear fit, the GCTA-DFT kinetic value at the TDL is certainly not correct (due to different charges).
However, it is interesting that the CTA-DFT potential energy TDL estimate, which is the same as the GCTA-SAFL value within the error bars, shows better linear fits despite the shell-filling effects not correctly captured in CTA-DFT.
This might be specific to NM systems since we observed a slightly different behavior in the FM $\alpha$-Fe.

\begin{table}[!htbp]
\centering
\caption{
Quality of linear fits using least squares regression in the TDL extrapolation of DMC energy components of Al.
Tabulated metrics are $R$-Squared ($R^2$, unitless, ideally 1) and the standard deviation of the slope ($\sigma_\mathrm{S}$, [Ha] units, ideally 0). 
The values are shown for DMC total, kinetic, and potential energies.
The least error quantities are indicated by bold text.
}
\label{tab:Al_TDL_fit}
\begin{tabular}{cccccccc}
\hline
Energy & Metric              &    CTA-DFT  &  GCTA-DFT &  GCTA-SAFL \\
\hline
Tot.   & $\sigma_\mathrm{S}$ &    0.023478 &  0.001322 &  \textbf{0.000559} \\
Tot.   & $R^2$               &    0.993144 &  0.999992 &  \textbf{0.999999} \\
\\
Kin.   & $\sigma_\mathrm{S}$ &    0.038311 &  \textbf{0.010557} &  0.017360 \\
Kin.   & $R^2$               &    0.744534 &  \textbf{0.995669} &  0.992373 \\
\\
Pot.   & $\sigma_\mathrm{S}$ &    \textbf{0.007506} &  0.010240 &  0.018059 \\
Pot.   & $R^2$               &    \textbf{0.999071} &  0.998231 &  0.996741 \\
\hline
\end{tabular}
\end{table}

To validate that GCTA-SAFL is converging to the correct TDL total energy, we now compare the DMC cohesive energy ($E_\mathrm{coh}$) with those of experiments \cite{chase_jr_janaf_1985} and other many-body methods.
Table \ref{tab:Al_coh} reports the Al cohesive energies obtained from various theories.
The many-body methods we reference are a previous DMC calculation with canonical TA and coupled cluster with single and double excitations (CCSD); see Ref. \cite{neufeld_ground-state_2022} for explanations of additional corrections (CCSD(T)$_\mathrm{SR}$ and CCSD-SVC).
First, we see that LDA overestimates, in qualitative agreement with previous studies \cite{hood_diffusion_2012, gaudoin_ab_2002}, while HF severely underestimates the $E_\mathrm{coh}$ \cite{neufeld_ground-state_2022}.
Using CCSD and further corrections brings the value closer to the experimental value.
The previous DMC calculation with a backflow transformed wave function (BF) \cite{lopez_rios_inhomogeneous_2006} and canonical TA is only off by 29 meV. 
Finally, DMC with GCTA-SAFL occupations and LDA orbitals show an excellent agreement with the experimental value.
We do not expect a severe sensitivity of DMC total energy to the underlying DFT exchange-correlation functional due to the simple, closed-shell NM phase of Al.
This exceptional agreement confirms that not only does GCTA-SAFL have a near-ideal linear behavior, but it also approaches the right asymptotic limit.

Let us comment on the difference in $E\rm^{Al}_{coh}$ between this work and the DMC/SJ value of Ref. \cite{hood_diffusion_2012} shown in Table \ref{tab:Al_coh}.
This difference of $\approx 0.1$~eV is most likely due to a combination of many differences in methodologies, such as different pseudopotentials (ccECP vs. Troullier-Martins \cite{troullier_efficient_1991}), slightly different geometries [$a = 4.0317(2)$~\AA~ vs. $a = 4.030(1)$~\AA], different underlying functionals (LDA vs. PBE \cite{perdew_generalized_1996}), and different nonlocal pseudopotential treatments (T-moves vs. locality approximation \cite{mitas_nonlocal_1991}), where the values in parentheses correspond to this work vs. Ref~\cite{hood_diffusion_2012} respectively.
Here, the difference due to GCTA vs. CTA of Ref.~\cite{hood_diffusion_2012} is expected to be minor, as they have used very large supercells (up to 1331 atoms) and excluded the small cells to eliminate the biases of CTA.
On the other hand, this work used small to moderately large supercells (up to 108 atoms) and obtained high-accuracy extrapolations.
This is an advantage of using GCTA since even the small cells can be used in extrapolations due to the absence of CTA biases that stem from occupying the states above the Fermi level or leaving the states empty below the Fermi level.

Another insight from Table~\ref{tab:Al_coh} is that the role of BF wave functions in obtaining the cohesive energy of Al may not be as important as previously thought.
The single-reference wave function already shows a good cancellation of errors, which was also observed to be the case in the main-group elemental Si solid \cite{annaberdiyev_cohesion_2021}.
Specifically, the SJ wave function in Si recovered more than $98\%$ of the correlation energy.
Therefore, provided that pseudopotentials are reliable and metallic occupations are properly set using GCTA, it is not unexpected to see accurate results with single-reference wave functions in this material.
Nevertheless, an explicit study using BF, GCTA, and accurate pseudopotentials would be beneficial to fully testing the accuracy of the SJ wave function.

\begin{table}
\centering
\caption{Cohesive energies [eV] of Al predicted by various methods.
The theoretical values do not include the zero-point-energy (ZPE) contributions.
The experimentally reported number ($327.320 \pm 4.2$ kJ/mol) was corrected for ZPE (0.04~eV) \cite{zhang_performance_2018} for consistency with the theory.
}
\begin{tabular}{llc}
    \hline
    Method                & $E_\mathrm{coh}$ [eV] & Ref. \\
    \hline
    LDA                   &    3.884  & this work \\
    \\
    HF                    &    1.388  & Ref.\cite{neufeld_ground-state_2022} \\
    CCSD                  &    2.966  & Ref.\cite{neufeld_ground-state_2022} \\
    CCSD(T)$_\mathrm{SR}$ &    3.102  & Ref.\cite{neufeld_ground-state_2022} \\
    CCSD-SVC              &    3.347  & Ref.\cite{neufeld_ground-state_2022} \\
    \\
    DMC/SJ                &  3.341(1) & Ref.\cite{hood_diffusion_2012} \\
    DMC/BF                &  3.403(1) & Ref.\cite{hood_diffusion_2012} \\
    DMC/SJ                &  3.438(2) & this work \\
    \\
    Experiment            & 3.432(44) & Ref.\cite{chase_jr_janaf_1985} \\
    \hline
\end{tabular}
\label{tab:Al_coh}
\end{table}

\subsubsection{Fe}

\begin{figure*}[!htbp]
\centering
\begin{subfigure}{0.33\textwidth}
\includegraphics[width=\textwidth]{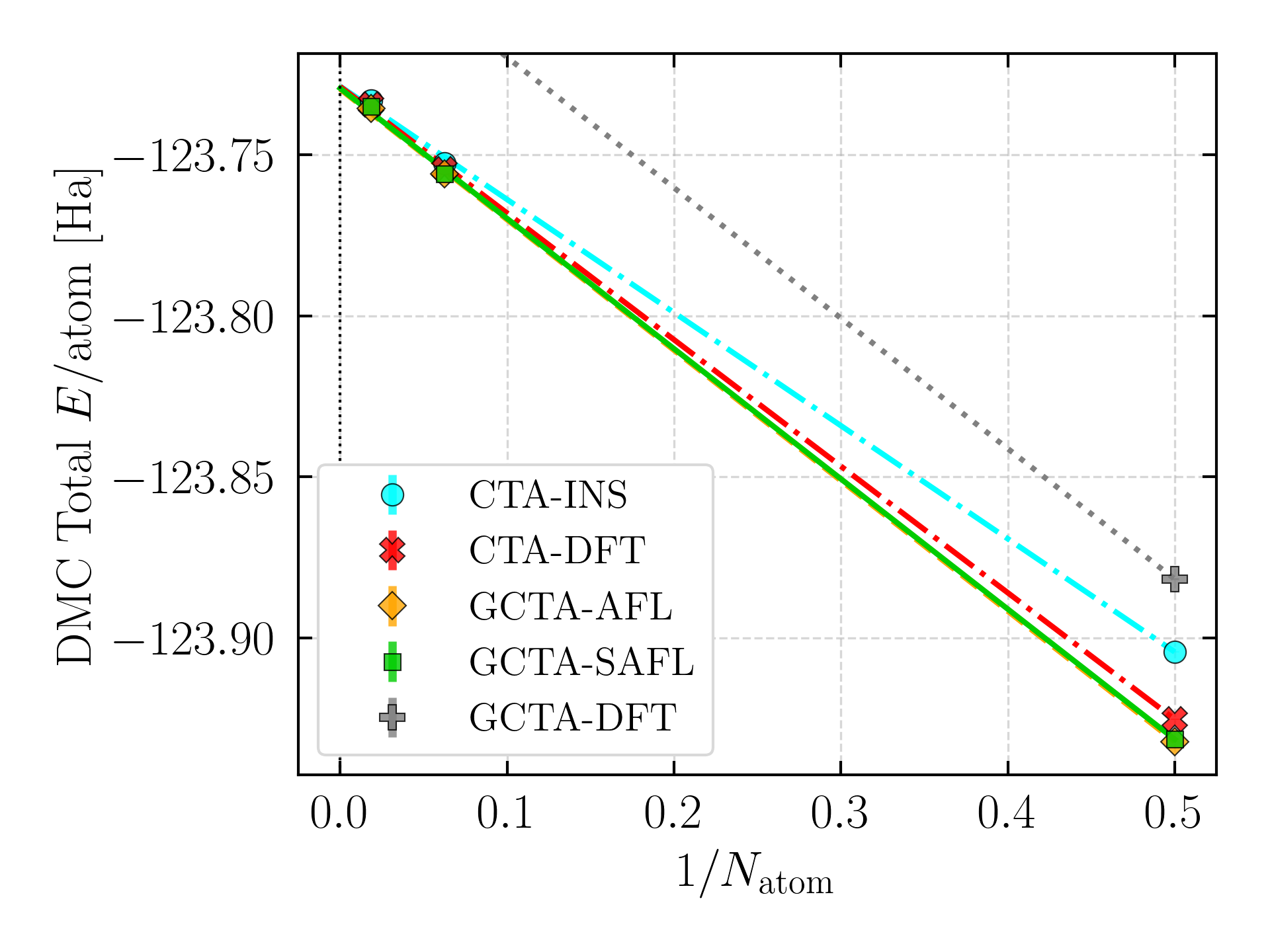}
\caption{$E$/atom}
\label{fig:Fe_FM_TDL_dmc-tot}
\end{subfigure}%
\begin{subfigure}{0.33\textwidth}
\includegraphics[width=\textwidth]{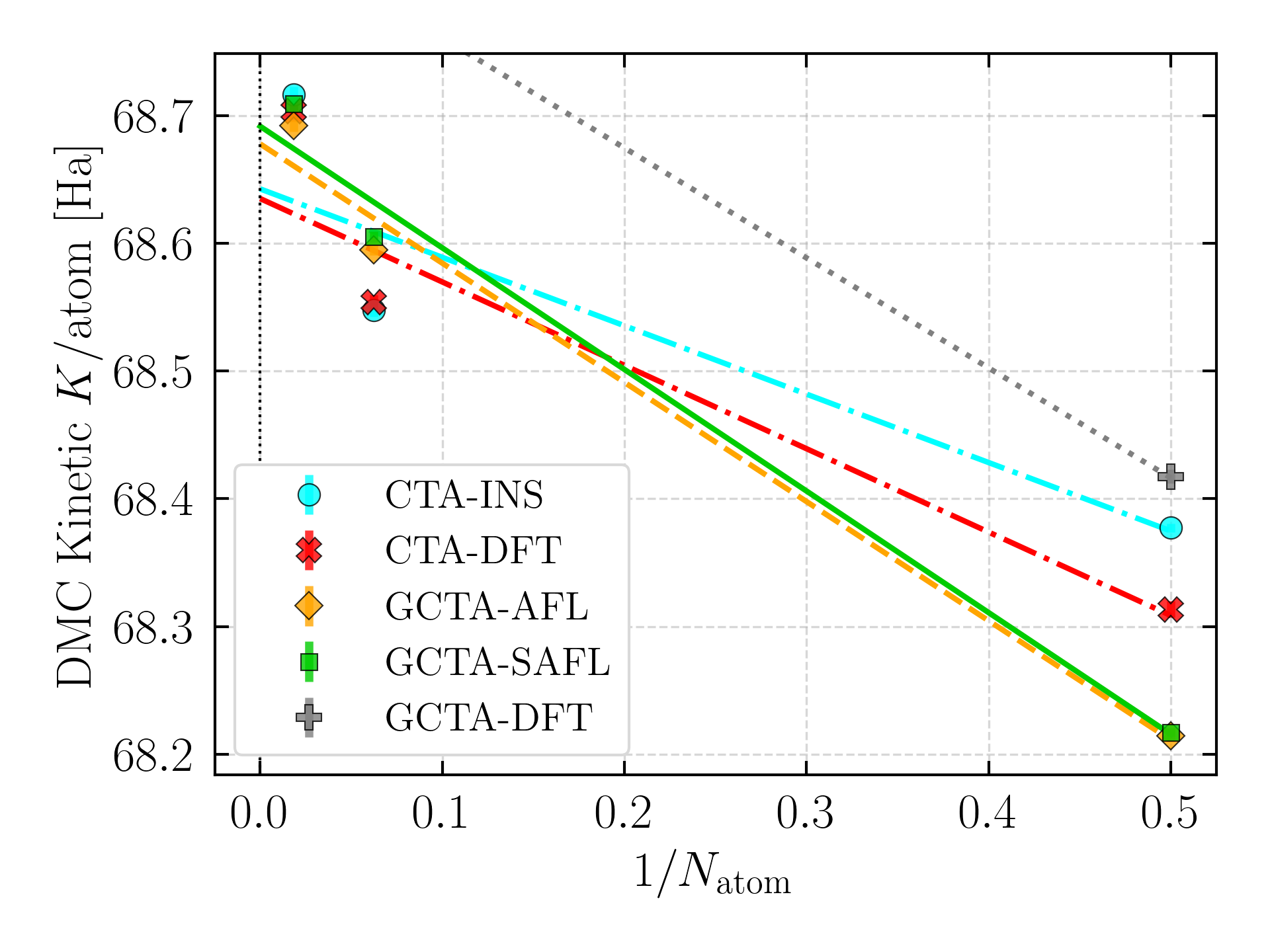}
\caption{$K$/atom}
\label{fig:Fe_FM_TDL_dmc-kin}
\end{subfigure}%
\begin{subfigure}{0.33\textwidth}
\includegraphics[width=\textwidth]{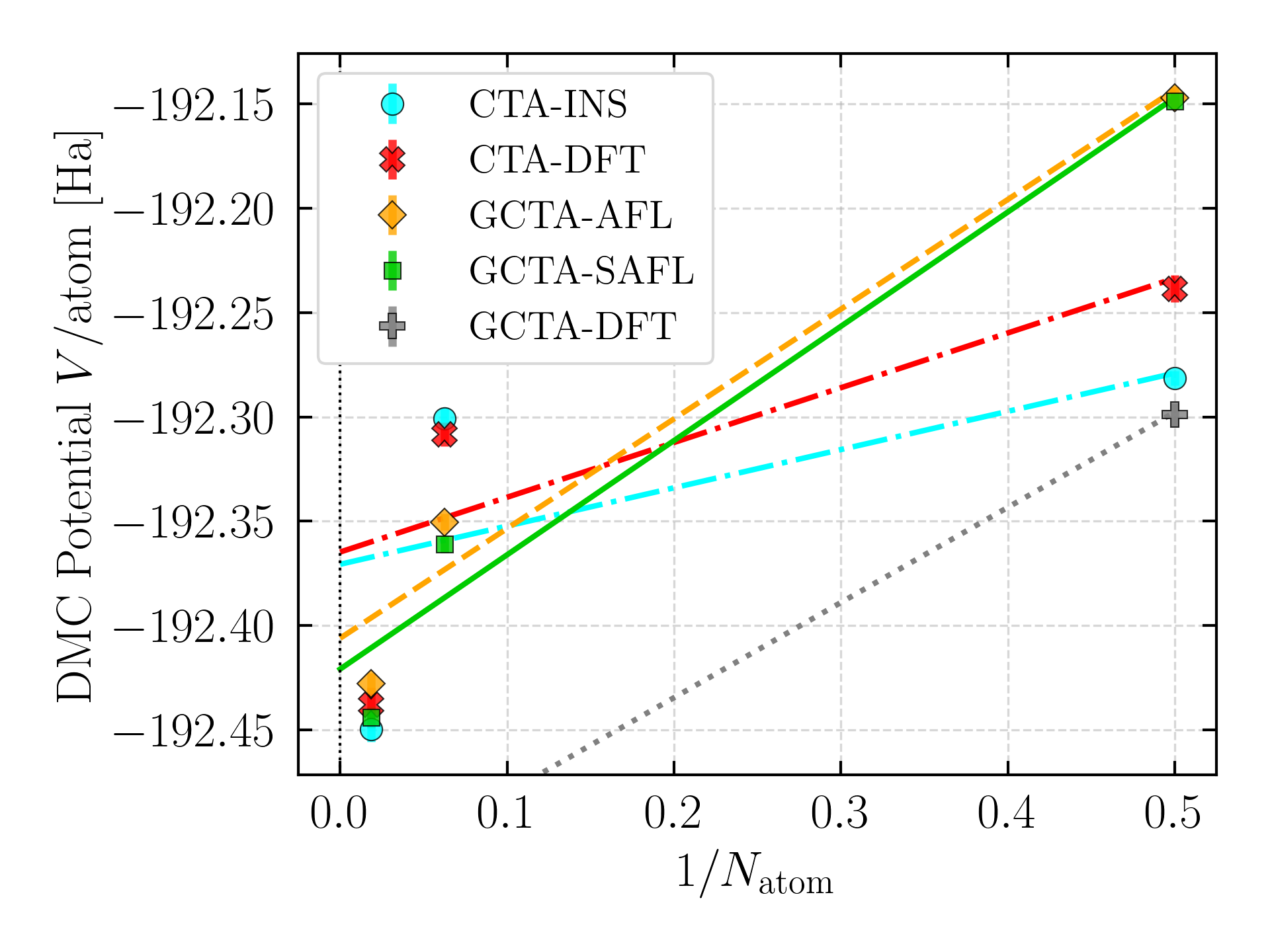}
\caption{$V$/atom}
\label{fig:Fe_FM_TDL_dmc-pot}
\end{subfigure}
\begin{subfigure}{0.33\textwidth}
\includegraphics[width=\textwidth]{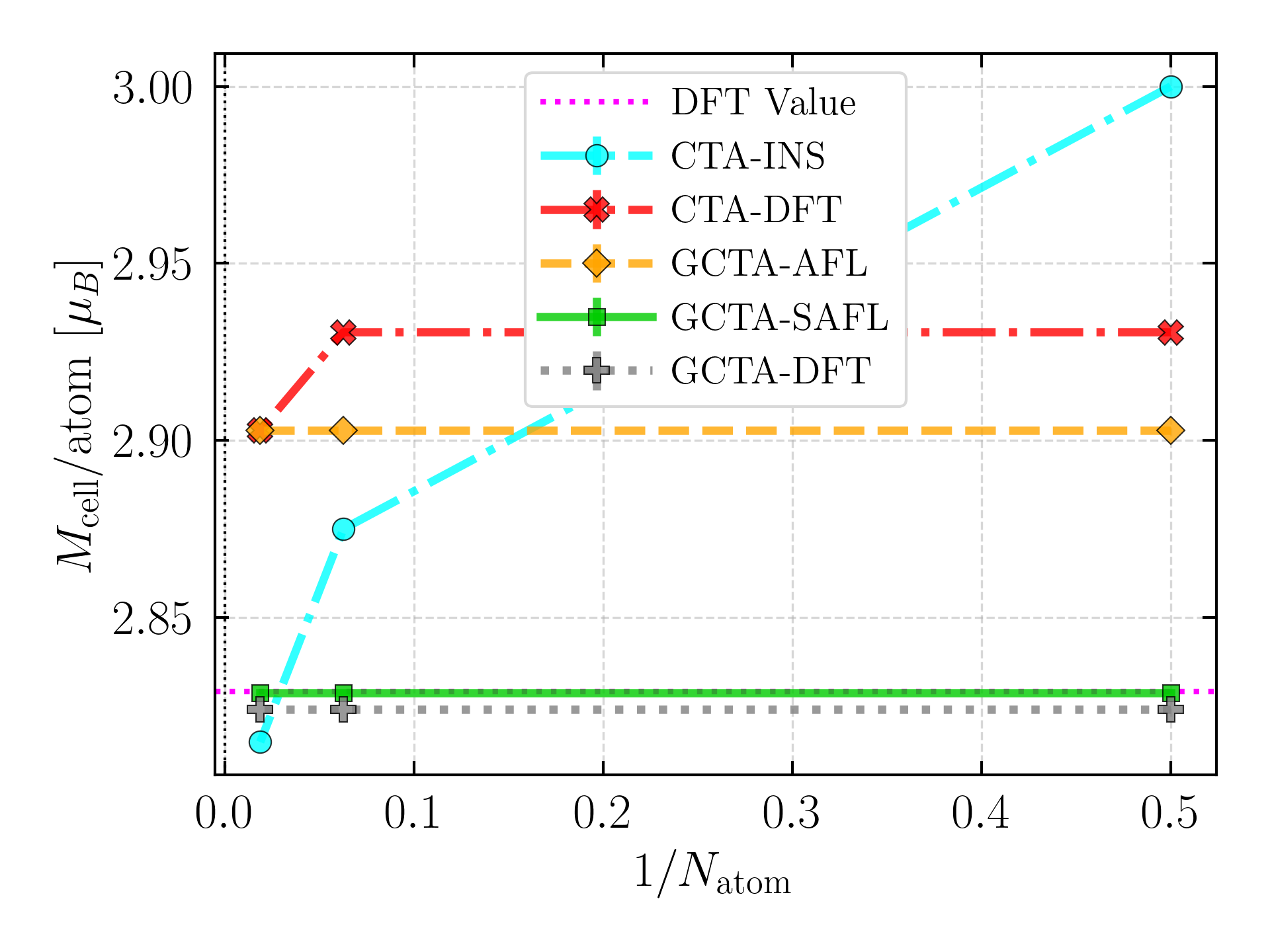}
\caption{$M_\mathrm{cell}$/atom}
\label{fig:Fe_FM_TDL_magnetization}
\end{subfigure}%
\begin{subfigure}{0.33\textwidth}
\includegraphics[width=\textwidth]{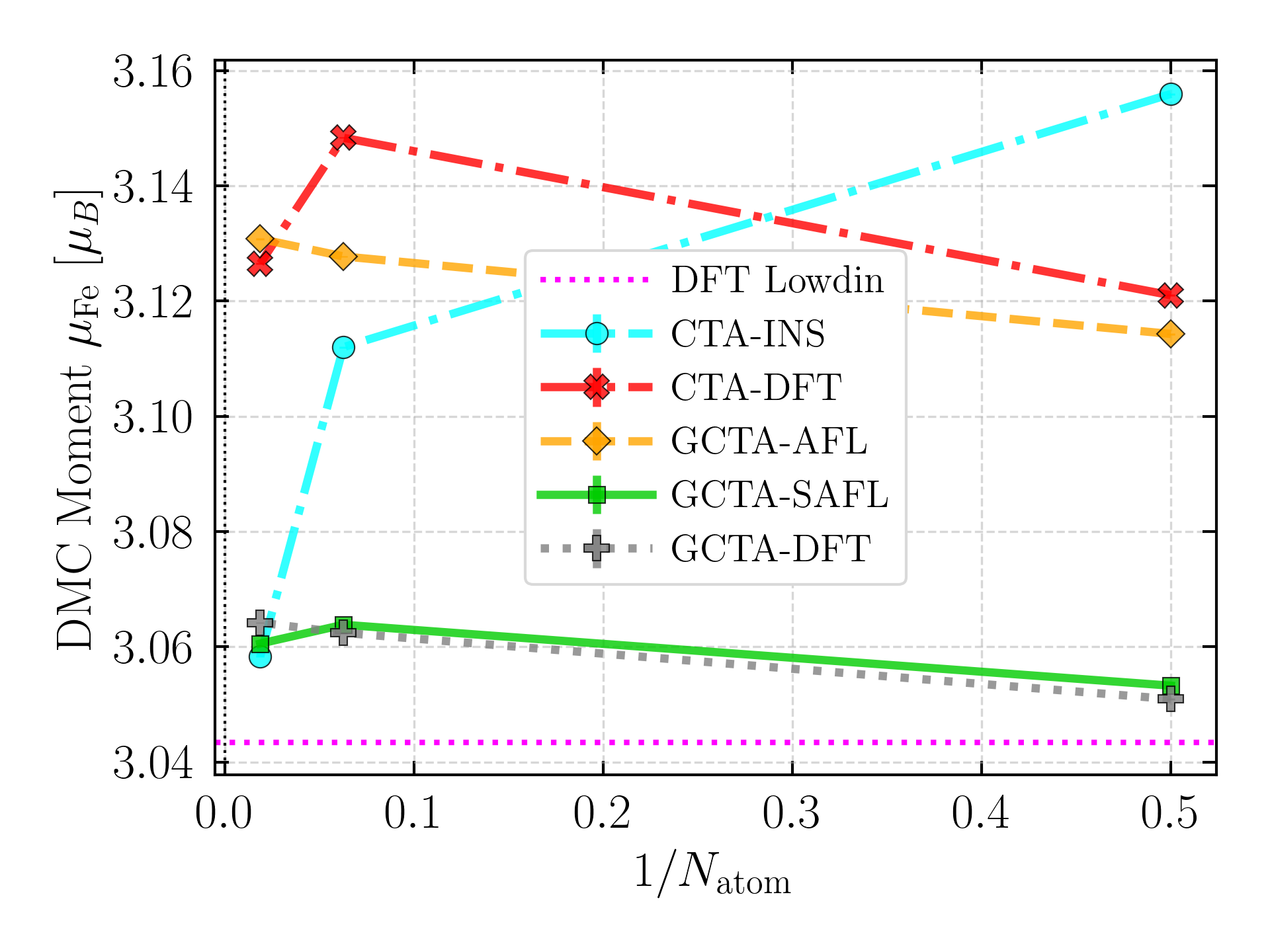}
\caption{$\mu_\mathrm{Fe}$}
\label{fig:Fe_FM_TDL_dmc-mom}
\end{subfigure}%
\begin{subfigure}{0.33\textwidth}
\includegraphics[width=\textwidth]{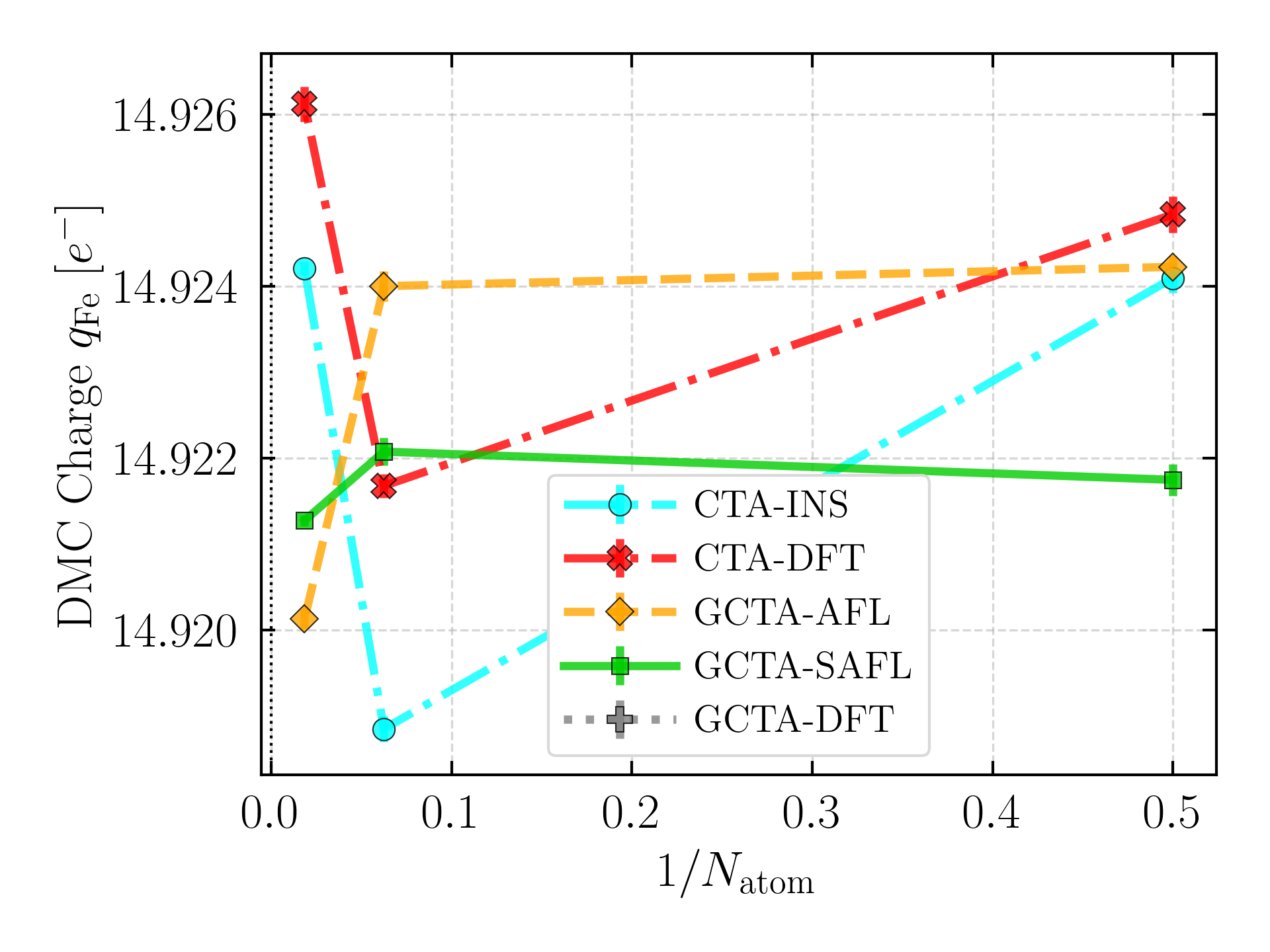}
\caption{$q_\mathrm{Fe}$}
\label{fig:Fe_FM_TDL_dmc-chg}
\end{subfigure}
\caption{
TDL convergence of twist-averaged DMC quantities of $\alpha$-Fe as the supercell is expanded.
(a) DMC total energy,
(b) DMC kinetic energy,
(c) DMC potential energy,
(d) cell magnetization,
(e) DMC atomic moment, and
(f) DMC atomic charge.
In (f), GCTA-DFT is outside of the visible window.
$Z_T\cdot Z_\theta = 6^3$ is kept constant as the supercell is increased. 
QMC errors are one standard deviation and smaller than the data symbol sizes in most cases.
}
\label{fig:Fe_FM_TDL_conv}
\end{figure*}

Figure \ref{fig:Fe_FM_TDL_conv} shows the DMC convergence for energies and magnetic properties of $\alpha$-Fe.
The overall picture for total, kinetic, and potential energy convergence is similar to that in the case of Al.
Here, the total energy extrapolations (Figure \ref{fig:Fe_FM_TDL_dmc-tot}) for all occupation schemes result in similarly accurate fits, with $R^2 > 0.999$ in all cases.
More pronounced differences can be seen in the kinetic (Figure \ref{fig:Fe_FM_TDL_dmc-kin}) and potential (Figure \ref{fig:Fe_FM_TDL_dmc-pot}) energy components.
The CTA schemes display large, sudden changes in the slopes.
However, the GCTA-SAFL and GCTA-AFL data are much closer to linear behavior.

Table \ref{tab:Fe_TDL_fit} provides the metrics for the quality of linear fits.
Here, the least errors occur for GCTA-DFT and GCTA-AFL.
This might be explained by GCTA-DFT and GCTA-AFL not having independent artificial shifts in $E_\mathrm{F}$ for up and down spin channels.
However, note that GCTA-DFT again extrapolates to incorrect TDL values and GCTA-SAFL fit quality is very close to GCTA-AFL.
In fact, the $\sigma_\mathrm{S}$ values of kinetic and potential energies of GCTA-SAFL are better than those of GCTA-AFL.
Also, GCTA-AFL results in slower convergence of magnetic properties, as shown in Figures \ref{fig:Fe_FM_T1_k_magnetization} and \ref{fig:Fe_FM_T1_k_dmc-mom}.
The important difference is that GCTA-AFL introduces an \textit{uncontrolled} deviation from the reference magnetization.
For example, the cell magnetization $M_\mathrm{cell}$ in Figure \ref{fig:Fe_FM_TDL_magnetization} shows that a constant shift persists to TDL in GCTA-AFL when using $Z_T \cdot Z_\theta = 6^3$ (although it should eventually approach the correct $M_\mathrm{cell}$ for large enough $Z_T \cdot Z_\theta$; see the inset of Figure \ref{fig:Fe_FM_T1_k_magnetization}).
GCTA-SAFL, on the other hand, reproduces the target magnetization (dotted horizontal line) within a $0.001~\mu_\mathrm{B}$.
GCTA-DFT is also very close and within $0.01~\mu_\mathrm{B}$ of target magnetization.
CTA-INS approaches the right value for the largest supercell since the target magnetization is included in the recipe but can only be improved by increasing the supercell.

The trends in the Fe magnetic moments (Figure \ref{fig:Fe_FM_TDL_dmc-mom}) are similar to those of the cell magnetization (Figure \ref{fig:Fe_FM_TDL_magnetization}).
However, all GCTA methods show a slight increase in the atomic moments as the supercell increases, despite the constant $M_\mathrm{cell}$.
This can be seen as the impact of many-body effects and interactions between different folded $k$-points, which are not captured in the primitive cell.
Finally, Figure \ref{fig:Fe_FM_TDL_dmc-chg} shows the Fe atomic charges.
All cases show small fluctuations around the same value of $14.92~e^{-}$, except for GCTA-DFT, which is not visible on this scale.

\begin{table*}[!htbp]
\centering
\caption{
Quality of linear fits using least squares regression in the TDL extrapolation of DMC energy components of $\alpha$-Fe.
Tabulated metrics are $R$-Squared ($R^2$, unitless, ideally 1) and the standard deviation of the slope ($\sigma_\mathrm{S}$, [Ha] units, ideally 0). 
The values are shown for DMC total, kinetic, and potential energies.
The least error quantities are indicated by bold text.
}
\label{tab:Fe_TDL_fit}
\begin{tabular}{lccccccc}
\hline
Energy   & Metric              & CTA-INS & CTA-DFT & GCTA-DFT & GCTA-AFL & GCTA-SAFL \\
\hline
Total     & $\sigma_\mathrm{S}$ & 0.00631 &  0.00391 & \textbf{0.00252} & 0.00325 & 0.00567 \\
Total     & $R^2$               & 0.99949 &  0.99982 & \textbf{0.99990} & 0.99988 & 0.99978 \\
\\
Kinetic   & $\sigma_\mathrm{S}$ & 0.20683 &  0.21320 & \textbf{0.06958} & 0.11576 & 0.09767 \\
Kinetic   & $R^2$               & 0.81242 &  0.89485 & 0.98619 & \textbf{0.98721} & 0.98541 \\
\\
Potential & $\sigma_\mathrm{S}$ & 0.20105 &  0.21093 & \textbf{0.06794} & 0.11725 & 0.09209 \\
Potential & $R^2$               & 0.39683 &  0.62348 & 0.95769 & \textbf{0.96385} & 0.96191 \\
\hline
\end{tabular}
\end{table*}

\section{Discussion}
\label{sec:discussion}

This work compared the convergence behavior of five different occupation schemes, although this list is not exhaustive.
For example, we have not included the GCTA sampling of the grand potential \cite{azadi_efficient_2019}.
Similarly, single twist sampling methods such as $\Gamma$-point sampling (PBC), Baldereschi point \cite{baldereschi_mean-value_1973}, or ``exact special twist" (EST) \cite{dagrada_exact_2016} point sampling were not included since the cell magnetization for these will have to be integers as in the case of CTA-INS.
This makes the supercell extrapolations (Figure \ref{fig:Fe_FM_TDL_conv}) challenging as the magnetic phase will evolve with the supercell.
However, we note that the EST method shows good energy convergence for the NM phase (and expectedly for the AFM phase).
The EST method might be particularly beneficial in insulators for optical or fundamental gap calculations, which require single-particle excitation at a particular twist.

Despite not considering the above methods, we demonstrated that GCTA-(S)AFL observables are already accurate enough for practical QMC applications with small or moderate sizes of twist-mesh.
For example, both schemes can be readily used to study metal-insulator transition \cite{bennett_new_2018} as they naturally reproduce CTA occupations for semiconductors or insulators.
Another promising avenue is using GCTA-(S)AFL in equation-of-state calculations where the lattice parameter is expanded or contracted \cite{hood_diffusion_2012}.
These calculations are challenging for CTA with small cells since Fermi level band crossings may occur as the lattice parameter is changed, which would not be appropriately accounted for by CTA, while this is not an issue for GCTA-(S)AFL.
Another advantage of GCTA over CTA methods is the faster convergence of the Fermi surface \cite{lin_twist-averaged_2001}.
However, all presented TA schemes, including the single twist methods, should approach the same energy, magnetism, and Fermi surface in the thermodynamic limit provided that charge neutrality is met.


It is important to note that, although not formally formulated and benchmarked, GCTA-AFL type occupation has been used before.
For example, Ref. \cite{al-hamdani_unraveling_2023} studied hydrogen adsorption using GCTA with $[3 \times 3 \times 1]$ $k$-grid and manually set the occupations close to the DFT values within the constraint of charge neutrality.
Here we have formulated, generalized, and automated this approach for any $k$-grid and magnetic systems, while also providing benchmarks for various TA schemes.
Automated workflows will facilitate the seamless integration of QMC methods into high-throughput calculations, eliminating the need for human input.

A source of possible complication with GCTA-AFL, as noted in the aforementioned Ref. \cite{al-hamdani_unraveling_2023} is a single-particle degeneracy in the eigenvalue sets (Equations \ref{eqn:eig_afl_set}, \ref{eqn:safl_set_up}, \ref{eqn:safl_set_dn}) near the Fermi level.
This type of degeneracy might produce a system with a nonzero net charge, particularly evident in coarse twist-mesh sizes, as in $[3 \times 3 \times 1]$ $k$-grid of Ca decorated graphene \cite{al-hamdani_unraveling_2023} or $[2 \times 2 \times 2]$ $k$-grid of Al \cite{gaudoin_ab_2002}.
In practice, for the $[2 \times 2 \times 2]$ $k$-grid of Al, we have observed that this issue is not significant since the eigenvalues are not numerically identical, and GCTA-AFL will occupy one (or a few) of the nearly degenerate states.
Additionally, the use of dense twist-mesh sizes in production calculations reduces the likelihood of eigenvalue degeneracies near Fermi level.
Furthermore, introducing various symmetry-breaking techniques widely employed in DFT \cite{malyi_false_2020} can further mitigate this issue.

We employed a uniform $k$-point grid throughout this work.
In the context of such a grid, GCTA-(S)AFL is equivalent to folding all $k$-points to form a large supercell with ($Z_T\cdot Z_\theta$) primitive cells and neutrally occupying the $\Gamma$-point to determine the appropriate Fermi level.
Subsequently, the adopted Fermi level can be used to unfold back to the primitive cell to obtain the necessary occupations.
However, a uniform grid is not a requirement for GCTA-(S)AFL, and any type of $k$-point choice, such as pseudorandom or quasi-random selections, can be readily employed.
Nevertheless, we recommend using GCTA-(S)AFL with Baldereschi-centered (or EST-centered) Monkhorst-Pack $k$-mesh, which was shown to display superior convergence than $\Gamma$-centered or random selection of $k$-points \cite{drummond_finite-size_2008, azadi_systematic_2015, dagrada_exact_2016}.

GCTA-SAFL is applicable only in collinear spin systems.
For noncollinear calculations, GCTA-AFL is a natural option that is applicable without any modifications since it only considers the lowest eigenvalues.
The introduced small deviation from the reference $M_\mathrm{cell}$ can be kept constant by keeping $Z_T \cdot Z_\theta$ unchanged during TDL extrapolation.
Thus, the QMC energies would correspond to the magnetism obtained within GCTA-AFL and not the reference magnetization (although the difference is quite small for a large twist-mesh).

Although we have shown only DMC results, we expect similar improvements in other real-space QMC flavors, such as variational Monte Carlo (VMC), or in second-quantized orbital-based flavors, such as auxiliary-field quantum Monte Carlo (AFQMC)~\cite{zhang_auxiliary-field_2018, qin_benchmark_2016, zhang_quantum_2003, purwanto_quantum_2004}.
In addition, since GCTA-(S)AFL only changes the occupation numbers in each twist, beyond single-reference wave functions such as backflow or multireference configuration-interaction (CI) expansions can be used as usual to improve the results further.

We have not attempted to calculate the cohesive energy of $\alpha$-Fe in this work.
This is because a large Hubbard $U$ was chosen to amplify the magnetic moments so that greater differences can be seen between various TA schemes.
Namely, unlike in Al, the Hubbard $U$ also changes the magnetic phase, and the trial wave function nodal surface is quite sensitive to this change \cite{kolorenc_wave_2010, townsend_starting-point-independent_2020, annaberdiyev_role_2023-1}.
Therefore, an accurate study must also predict a proper Hubbard $U$ or hybrid exact exchange fraction.
This is not the focus of this work, and we leave this important aspect for a future study.

\section{Conclusions}
\label{sec:conclusions}

This work proposed new grand-canonical twist-averaging with (spin)-adapted Fermi levels (GCTA-(S)AFL) as improved versions of the traditionally used twist-averaging schemes.
GCTA-SAFL shows the best overall convergence to the thermodynamic limit when considering the total energies and magnetization.
It shows a rapid convergence of energy components and magnetic moments as the number of twists is increased for a given simulation cell.
GCTA-SAFL and GCTA-AFL obtain lower energies than canonical twist-averaging methods for a large number of twists and show robust linear convergence in supercell extrapolations, allowing for thermodynamic limit estimations using smaller cells.
The proposed change is simple and only requires the availability of the single-particle orbital eigenvalues for each twist and light preprocessing to set the up and down occupations in the quantum Monte Carlo (QMC) code input.
Another welcome feature is that these schemes do not require any postprocessing corrections.
The resulting total energies can be directly compared to canonical twist-averaging values due to guaranteed charge neutrality, and the extrapolated values find the correct asymptotic limit, as corroborated by the Al experimental value.

The proposed GCTA-(S)AFL eliminates the uncertainties of setting the occupations in magnetic metals within the QMC methods.
This, coupled with reductions in the number of $k$-points due to Brillouin zone symmetry, enables accurate modeling of magnetic metals within QMC.
The resolution to this technical aspect will allow for studies of more central challenges, such as proper (de)localization and multireference effects in magnetic metals.
Therefore, the proposed improvement provides a vital building block in bridging the gap between QMC calculations and experimental observables for metallic materials.

\section*{Associated Content}
The input and output files generated in this work were published in the Materials Data Facility \cite{blaiszik_materials_2016, blaiszik_data_2019} and can be found in Ref.~\cite{mdf_data}.
See the SI for the employed bulk geometry, extended DFT data, and extended QMC data.
This information is available free of charge via the Internet at \url{http://pubs.acs.org}.

\begin{acknowledgments}

We thank P. R. C. Kent and K. Saritas for reading the manuscript and providing helpful suggestions.

This work has been supported by the U.S. Department of Energy, Office of Science, Basic Energy Sciences, Materials Sciences and Engineering Division, as part of the Computational Materials Sciences Program and Center for Predictive Simulation of Functional Materials.
An award of computer time was provided by the Innovative and Novel Computational Impact on Theory and Experiment (INCITE) program.
This research used resources of the Argonne Leadership
Computing Facility, which is a DOE Office of Science User Facility supported under Contract No. DE-AC02-06CH11357.
This research used resources of the Oak Ridge Leadership Computing Facility, which is a DOE Office of Science User Facility supported under Contract No. DE-AC05-00OR22725.
This research used resources of the National Energy Research Scientific Computing Center (NERSC), a U.S. Department of Energy Office of Science User Facility located at Lawrence Berkeley National Laboratory, operated under Contract No. DE-AC02-05CH11231.

This paper describes objective technical results and analysis. Any subjective views or opinions that might be expressed in the paper do not necessarily represent the views of the U.S. Department of Energy or the United States Government.

Notice:  This manuscript has been authored by UT-Battelle, LLC, under contract DE-AC05-00OR22725 with the US Department of Energy (DOE). The US government retains and the publisher, by accepting the article for publication, acknowledges that the US government retains a nonexclusive, paid-up, irrevocable, worldwide license to publish or reproduce the published form of this manuscript, or allow others to do so, for US government purposes. DOE will provide public access to these results of federally sponsored research in accordance with the DOE Public Access Plan (http://energy.gov/downloads/doe-public-access-plan).

\end{acknowledgments}

\section*{Competing Interests}
The authors declare no competing financial or non-financial interests.

\section*{Author Contributions}
A.A. carried out the calculations and partially wrote the manuscript.
J.T.K. and P.G. supervised the project and aided in the writing of the manuscript.

\bibliography{main.bib}

%

\end{document}